\documentclass[12pt]{iopart}

\usepackage{graphicx}
\usepackage[utf8]{inputenc}
\usepackage[T1]{fontenc}
\usepackage{mathrsfs}
\usepackage{latexsym,amsfonts}
\usepackage{bm}

\usepackage{xcolor}

\expandafter\let\csname equation*\endcsname\relax
\expandafter\let\csname endequation*\endcsname\relax

\usepackage{amsmath}
\usepackage{amssymb}

\allowdisplaybreaks

\newcommand{\md}{\mathrm{d}}

\newcommand{\To}{{T_{\text{o}}}}

\newcommand{\btheta}{\boldsymbol{\theta}}
\newcommand{\bzeta}{\boldsymbol{\zeta}}
\newcommand{\bxi}{\boldsymbol{\xi}}

\newcommand{\Ys}{{{}^{-s}Y}}
\newcommand{\Ytwo}{{{}^{-2}Y}}

\newcommand{\be}{\begin{equation}}
\newcommand{\ee}{\end{equation}}
\newcommand{\bea}{\begin{eqnarray}}
\newcommand{\eea}{\end{eqnarray}}

\newcommand{\av}[1]{{\left\langle#1\right\rangle}}

\eqnobysec

\begin{document}


\title[Search for Postmerger Gravitational Waves from Binary Neutron Star Mergers]
{Search for Postmerger Gravitational Waves from Binary Neutron Star Mergers
Using a~Matched-filtering Statistic}

\author{Andrzej Kr\'olak$^{1,2}$,
Piotr Jaranowski$^3$,
Michał Bejger$^{4,5}$,
Paweł Ciecieląg$^5$,
Orest Dorosh$^2$ and
Andrzej Pisarski$^3$}

\address{$^1$ Institute of Mathematics, Polish Academy of Sciences,
00-656 Warsaw, Poland}
\address{$^2$ National Center for Nuclear Research,
05-400 Świerk-Otwock, Poland}
\address{$^3$ Faculty of Physics, University of Bialystok, Ciolkowskiego 1L, 15-245 Bialystok, Poland}
\address{$^4$ INFN Sezione di Ferrara, Via Saragat 1, 44122 Ferrara, Italy}
\address{$^5$ Nicolaus Copernicus Astronomical Center,
Polish Academy of Sciences, 00-716, Warsaw, Poland} 

\ead{p.jaranowski@uwb.edu.pl}

\begin{abstract}

In this paper, we present a new method to search for a short, a few tens of milliseconds long,
postmerger gravitational-wave signal following the merger of two neutron stars.
Such a signal could follow the event GW170817 observed by LIGO and Virgo detectors. Our method is based
on a matched filtering statistic and an approximate template of the postmerger signal in the form of a damped sinusoid.
We test and validate our method using postmerger numerical simulations from
the CoRe database. We find no evidence of the
short postmerger signal in the LIGO data following the GW170817 event and we obtain upper limits.
For short postmerger signals investigated,
our best upper limit on the root sum square of the gravitational-wave strain emitted from 1.15 kHz to 4 kHz
is $h_{\text{rss}}^{50\%} = 1.8\times 10^{-22}/\sqrt{\text{Hz}}$ at 50\% detection efficiency.
The distance corresponding to this best upper limit is 4.64~Mpc.

\end{abstract} 


\date{\today}


\section{Introduction}
\label{sec:intro}

Gravitational-wave (GW) astronomy began dramatically with the discovery of 
GW150914 event on September 14, 2015,
a coalescence of two stellar-mass black holes (BHs) \cite{PhysRevLett.116.061102}.
Ninety more mergers of compact binaries involving black holes and neutron stars
were observed collectively in the O1 run of two Advanced Laser Interferometer Gravitational-Wave Observatory (LIGO)
detectors \cite{2015CQGra..32g4001L} and in the longer, more sensitive O2 and O3 runs,
in which the Advanced Virgo detector \cite{2015CQGra..32b4001A} joined the observations \cite{Abbott_2021}.
The most exciting was the GW170817 event, 
most likely the first detection of a binary neutron star (BNS) merger \cite{PhysRevLett.119.161101}.
Supporting this hypothesis were electromagnetic counterparts observed across the spectrum
\cite{Abbott_2017a,Abbott_2017b}. Thanks to its relatively close proximity to Earth, 
with 90\% credible interval of $40_{-14}^{+8}$~Mpc
for the distance measured by the GW data analysis \cite{PhysRevLett.119.161101},
GW170817 offers the first opportunity to study the nature of the remnant 
leftover from a BNS merger using GW observations.
Another likely BNS merger (event GW190425) was observed during the O3 run \cite{Abbott_2020}.
However, because of its large total mass of around 3.4 solar masses,
the remnant of GW190425 was expected to collapse promptly into a black hole.

The merger of two neutron stars (NSs) can have four possible outcomes:
(i) the prompt formation of a BH;
(ii) the formation of a hypermassive NS that collapses to a BH in $\lesssim1$~s;
(iii) the formation of a supramassive NS that collapses to a BH on timescales of $\sim$10--$10^4$~s;
or (iv) the formation of a stable NS.
The specific outcome of any merger depends on the progenitor masses
and on the NS equation of state (EOS).
The first detectability studies for the postmerger GW signals were performed 
in \cite{PhysRevD.90.062004,Clark_2016}.
In this paper, we introduce a matched-filtering approach
to search short ($\lesssim 1$~s) postmerger GW signal.
Our matched filters are based on an approximate model of postmerger signals
deduced from numerical relativity (NR) simulations.
We test our method using data from LIGO detectors
following the event GW170817.
As for the analysis of \cite{Abbott_2017}
that used the coherent Wave Burst (cWB) pipeline \cite{Klimenko-2008,PhysRevD.93.042004}
we find no evidence for a statistically significant signal
and we set upper limits on possible GW strain amplitudes.
See \cite{Putten_2018,Abbott-2019} for a search for longer-lived postmerger GWs following the GW170817 merger
and \cite{Thrane-2011,Miller-2018,Sun-Melatos-2019,Oliver-2019,Banagiri-2019,https://doi.org/10.1103/PhysRevD.100.062005} for search methods for long-duration GW transients.

This paper is organized as follows.
In section \ref{sec:sig} we present gravitational waveforms from mergers of two NSs
available in the Computational Relativity (CoRe)
database\footnote{{\tt http://www.computational-relativity.org}.} \cite{Dietrich_2018,CoRe2022}.
In section 3 we introduce an approximate analytic model of a postmerger waveform.
In section 4 we describe the response of a laser-interferometric detector to postmerger GWs.
In section 5 we introduce a matched-filtering statistic
to detect a postmerger signal in the noise of a network of detectors.
In section 6 we present the results of simulations consisting of adding waveforms from the CoRe database
to data from the network of LIGO detectors and testing their detectability with our match-filtering statistic
using approximate waveforms introduced in section 3.
In section 7 we present the search for postmerger signals
after the GW170817 event with the method developed in sections 3 and 5.
In section 8, using simulations from section 6, we obtain the sensitivity 
of our method to search for postmerger signals in the LIGO data after the GW170817 event.
In \ref{app:waveforms} we give details of derivations of postmerger waveforms.
In \ref{app:FAP} we present details of the calculation of the Fisher matrix and the false alarm
probability in application to searches for postmerger signals in the detector noise.

\section{Gravitational waveforms from binary neutron star merger}
\label{sec:sig}

In this paper, we investigate the detectability of the postmerger GWs
associated with the GW170817 event.
In the analysis, we use data collected by the LIGO detectors
and we employ waveform models,
which are available in the second release of
the CoRe numerical database \cite{CoRe2022}.

\begin{figure}
\includegraphics[width=\textwidth]{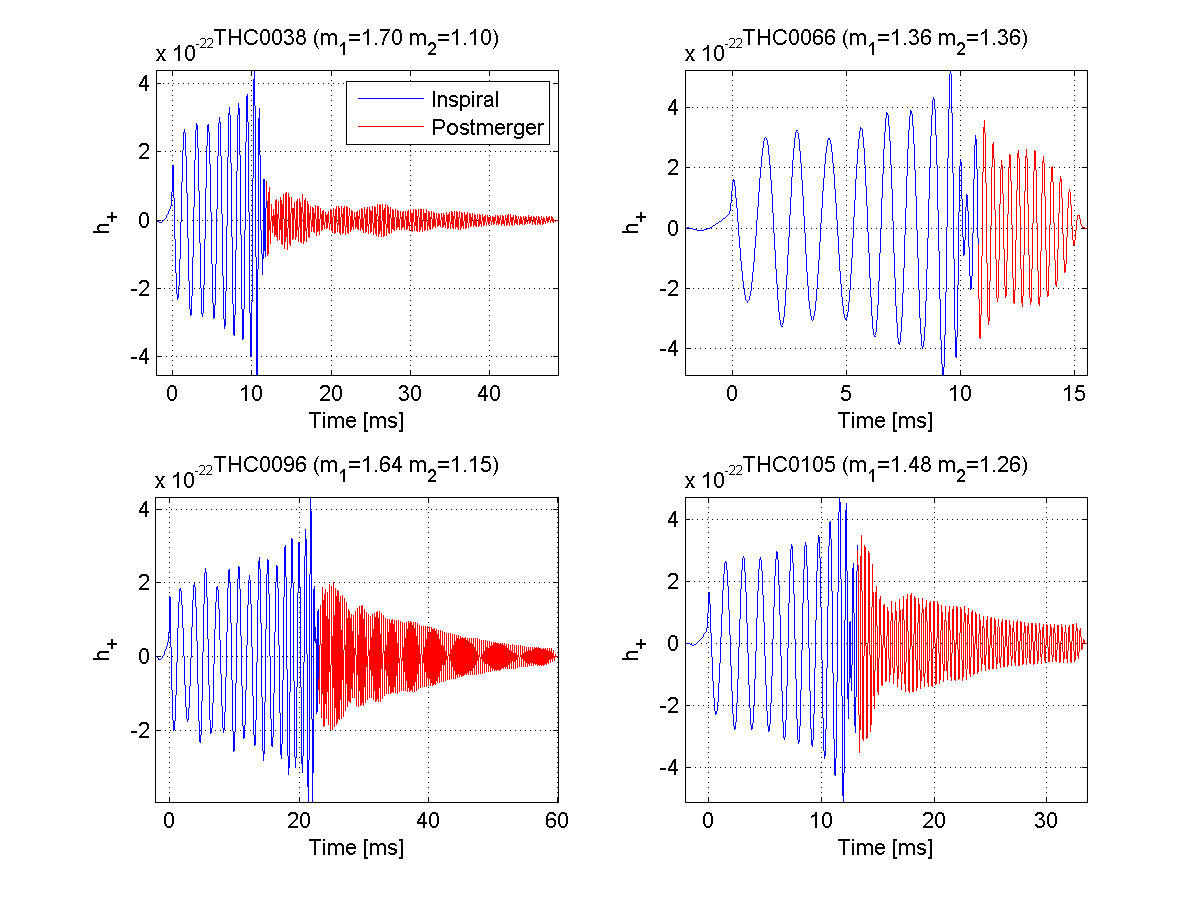}
\caption{Gravitational waveforms from binary neutron star mergers
calculated using data taken from the CoRe database.
The waveforms for four different numerical simulations are given;
the labels of the simulations are placed above the figures,
where we also give masses (in units of solar mass) of the components of the binary.
The inspiral phase of the waveforms is displayed in blue
and their postmerger part is marked in red.
Only the polarization $h_+$ including the dominant $l=2$, $m=\pm2$ modes is plotted,
$h_+$ is related to the amplitude and the phase of the $(2,2)$ mode by Eq.\ \eqref{eq:hpcmod2p}.
The waveforms were calculated
assuming that the distance to the source $r\cong40$~Mpc,
the inclination angle $\iota\cong156^\circ$,
and the angle $\phi=0^\circ$.}
\label{fig:waveforms_GW17}
\end{figure}

When looking for a signal, we restrict ourselves to waveforms consistent with the GW170817 event \cite{Nedora_2021}.
Four examples of such waveforms are presented in figure \ref{fig:waveforms_GW17}
(how to get these waveforms from the data directly downloaded from the CoRe database
is explained in \ref{app:waveforms}).
For our analysis, we consider only the dominant $l=2$ mode with  $m =\pm2$ components.
The postmerger signals (marked in red colour in figure \ref{fig:waveforms_GW17}) are defined
by the following criterion. Firstly we identify the merger time $t_m$ as the time
at which the function $h_+^2(t) + h_\times ^2(t)$ attains maximum.
Then we take for our analysis the signal that begins 1.25~ms after the merger time $t_m$.
In this way we discard the initial short-duration transients present in the postmerger signal.

\begin{figure}
\includegraphics[width=\textwidth]{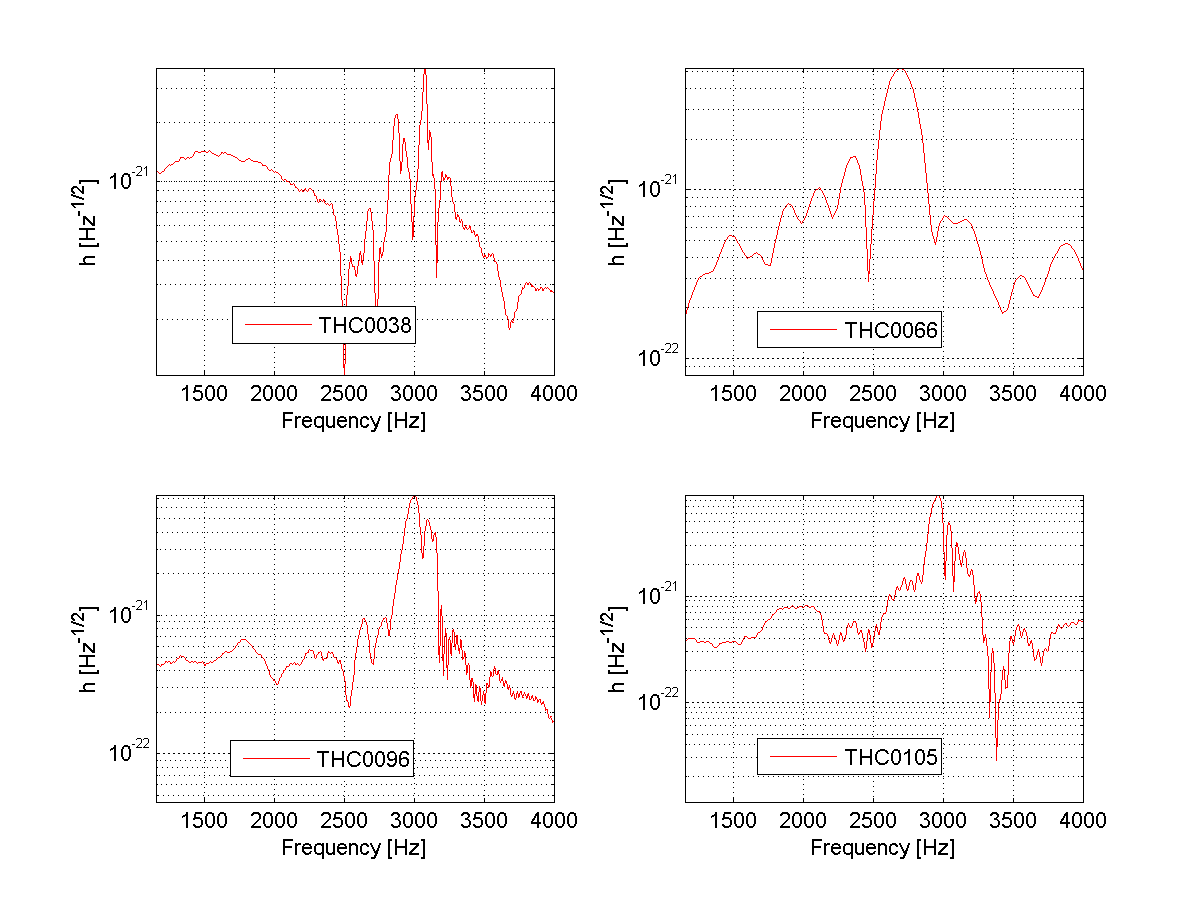}
\caption{Spectra of the postmerger waveforms (plotted in red in figure \ref{fig:waveforms_GW17}).
The power of the postmerger signal is concentrated around the maximum frequency $f_{\text{peak}}$ of the spectrum.
The spectra are obtained by passing the waveforms through a narrowband filter of
bandwidth $\langle1100;4100\rangle$~\text{Hz} and then performing the FFT.} 
\label{fig:spectra}
\end{figure}

In figure \ref{fig:spectra} we have plotted spectra of the signals depicted in figure \ref{fig:waveforms_GW17}.
We see that these spectra are dominated by a certain frequency $f_{\text{peak}}$
and the power of the signal is concentrated  around $f_{\text{peak}}$ in a band of a few hundred Hz wide.

\section{An approximate model of postmerger gravitational waves}
\label{sec:model}

In reference \cite{PhysRevLett.120.031102} the following approximate two-component analytical model
of the GW postmerger signal has been proposed (we consider here only the $+$ polarization of the GW
and assume that the signal begins at time $t=0$):
\be
h(t) = h_{short}(t) + h_{long}(t),
\ee
where the {\em short}-duration component is given by  
\begin{subequations}
\label{eq:pm_short}
\begin{align}
h_{short} &= \alpha A_1 \big[ \sin(2\pi f_1 t) + \sin(2\pi (f_1 - f_{1\epsilon}) t)
+ \sin(2\pi (f_1 + f_{1\epsilon})t) \big],
\\[1ex]
A_1 &= \exp(-t/\tau_1),
\end{align}
\end{subequations}
and the {\em long}-duration component reads
\begin{subequations}
\label{eq:pm_long}
\begin{align}
h_{long} &= A_2 \sin(2\pi f_2 t + 2\pi\gamma_2 t^2 + 2\pi\xi_2 t^3 + \pi \beta_2),
\\[1ex]
A_2 &= \exp(-t/\tau_2).
\end{align}
\end{subequations}
Here $f_{1\epsilon} = 50$~Hz and $f_1$, $\tau_1$, $f_2$, $\tau_2$
are characteristic frequencies and damping times
of the short and long component of the signal, respectively,
$\alpha$ gives the relative amplitude of the two components,
$\gamma_2$ and $\xi_2$ are parameters describing the evolution of the frequency $f_2$,
and the phase angle $\beta_2$ is adjusted to match numerical-relativity waveforms.
The fitted values of the above parameters are given in Table I of \cite{PhysRevLett.120.031102}
for several numerical postmerger signal simulations.

The short-duration component can be a combination tone between the
quadrupole and quasi-radial oscillations in the remnant or
a frequency due to tidally-formed orbiting bulges,
whereas the long component is just the quadrupole oscillation
with frequency $f_{\text{peak}}=f_2$, see \cite{PhysRevD.91.124056}.

\begin{figure}
\includegraphics[width=\textwidth]{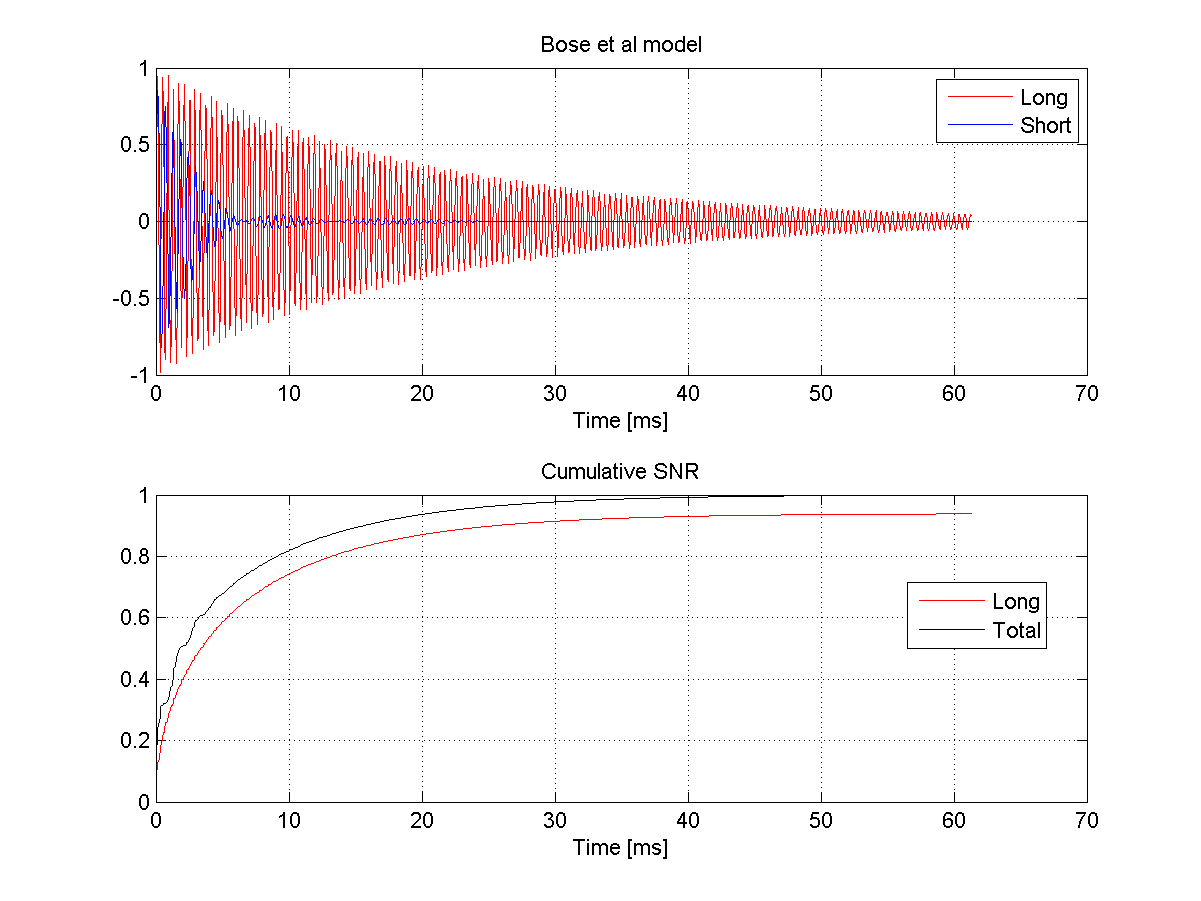}
\caption{Comparison of the {\em short} and {\em long} postmerger signals
given by Eqs.\ \eqref{eq:pm_short} and \eqref{eq:pm_long}
for the model H4-1325 with the fitted parameters taken from Table I of \cite{PhysRevLett.120.031102}.
The top panel shows the two waveforms and the bottom panel compares
the normalized cumulative SNR of the whole signal with the cumulative SNR of only the long signal.}
\label{fig:Bose_model}
\end{figure}

In figure \ref{fig:Bose_model} we have compared the importance
of both components of the postmerger signal
by displaying the normalized cumulative signal-to-noise ratios (SNRs) produced 
by the complete signal and by its long-duration component.
We see that the short-duration component contributes only around 5\%
to the overall SNR of the whole signal.
For this reason, we have neglected the short-duration component in our analysis.
The comprehensive tests of our approximation for 24 models from CoRe database
are given in Table 1, where in column 2 we give the SNR loss with respect to perfectly matched model
(we discuss these results at the end of section \ref{sec:injections}).

Consequently to search for the postmerger signal in the detector's noise
we propose the following approximate model of the $+$ and $\times$ GW polarizations:
\begin{subequations}
\label{hphc}
\begin{align}
h_+(t; A_{0+},\beta,o,\bzeta) &:= A_{0+}e^{-ot}\sin(\Phi(t;\bzeta)+\beta),
\\[1ex]
h_\times(t;A_{0\times},\beta,o,\bzeta) &:= A_{0\times}e^{-o t}\cos(\Phi(t;\bzeta)+\beta),
\end{align}
\end{subequations}
where the phase $\Phi$ reads
\be
\label{Phi}
\Phi(t;\bzeta) = 2\pi f t + 2\pi \gamma t^2.
\ee
Here $A_{0+}$ and $A_{0\times}$ are two constant amplitudes of the $+$ and $\times$ polarizations, respectively,
$o:=1/\tau$ is the inverse of the decay time $\tau$ (common for both waveforms),
and the vector $\bzeta$ collects the phase parameters,
\be
\bzeta := (f, \gamma).
\ee
Thus as our model we take the long component of the model from reference \cite{PhysRevLett.120.031102},
but with the $\propto t^3$ term in the phase discarded.

At the end of \ref{app:waveforms} we express the analytical model of
the $+$ and $\times$ GW polarizations defined by Eqs.\ \eqref{hphc} and \eqref{Phi}
in terms of the dominant $(2,\pm2)$ modes of the postmerger GWs.

Recently a more refined, 17-parameter model of the postmerger waveform
was obtained in \cite{https://doi.org/10.48550/arxiv.2205.09112}.
The dominant component of this model, denoted by $\tilde{W}_{\text{peak}}(t)$
(see figure 2 and Appendix C of \cite{https://doi.org/10.48550/arxiv.2205.09112})
is characterized by three parameters---frequency, frequency drift, and decay time,
similarly as in the model presented above. A similar, multiparameter analytical model has also
been proposed in \cite{PhysRevD.105.043020}.

\section{Response of the detector to postmerger gravitational waves}
\label{sec:response}

We assume that we have a network of $N$ laser-interferometric GW detectors.
The response function $s_I(t)$ of the $I$th detector to the postmerger GW has the form
\be
\label{eq:resI}
s_I(t) = \Theta(t)\Theta(\To-t)\big(F_{I+} h_+(t) + F_{I\times} h_\times(t)\big),
\ee
where $h_+(t)$ and $h_\times(t)$ are polarization functions of the GW,
$F_{I+}$ and $F_{I\times}$ are two beam pattern functions for the $I$th detector,
and $\Theta(t)$ is the step function (equal 0 for $t<0$ and 1 for $t\ge0$).
We thus conventionally assume that $t=0$ is the onset of the postmerger signal as seen by the $I$th detector,
and that $\langle0;\To\rangle$ is the observational interval.
To a very good accuracy, we can assume that the beam pattern functions $F_{I+}$ and $F_{I\times}$
are constant in the observational interval.
In our analysis,  we generate the responses \eqref{eq:resI} for the two LIGO detectors
using functions \verb"antenna_pattern" and \verb"project_wave" taken from
the module \verb"pycbc.detector" of a software package pyCBC \cite{alex_nitz_2023_7547919}.

After substituting Eq.\ \eqref{hphc} into Eq.\ \eqref{eq:resI},
the response function $s_I$ of the $I$th detector is given by
\be
\label{RF7}
s_I(t;A_{Is},A_{Ic},o,\bzeta) = \Theta(t)\Theta(\To-t)
\big(A_{Is} h_s(t;o,\bzeta) + A_{Ic} h_c(t;o,\bzeta)\big),
\ee
where the functions $h_s$ and $h_c$ read
\begin{subequations}
\label{RF8}
\begin{align}
h_s(t;o,\bzeta) &:= e^{-o t}\sin\Phi(t;\bzeta),
\\[1ex]
h_c(t;o,\bzeta) &:= e^{-o t}\cos\Phi(t;\bzeta).
\end{align}
\end{subequations}
The constants $A_{Is}$ and $A_{Ic}$ are different for different detectors
and they equal
\begin{subequations}
\begin{align}
A_{Is} &:= F_{I+}A_{0+}\cos\beta - F_{I\times}A_{0\times}\sin\beta,
\\[1ex]
A_{Ic} &:= F_{I+}A_{0+}\sin\beta + F_{I\times}A_{0\times}\cos\beta.
\end{align}
\end{subequations}

\section{Detection of the postmerger signal in the noise of the detector}
\label{sec:det}

In this section, we derive the detection statistic for the postmerger GW signal,
assuming that the noise in each detector is Gaussian, stationary, and a zero-mean stochastic process.
The derivation will be valid for more general signals than those defined by Eqs.\ \eqref{RF7}--\eqref{RF8}
and is similar to the derivation of the continuous-wave $\mathcal{F}$-statistic \cite{1998PhRvD..58f3001J,Cutler_Schutz_2005}.

We assume that the response function for the $I$th detector has the following form
\be
\label{gensig}
s_I(t;A_{Is},A_{Ic},\bxi) = \Theta(t)\Theta(\To-t)
\big(A_{Is} h_s(t;\bxi) + A_{Ic} h_c(t;\bxi)\big),
\ee
where $A_{Is}$, $A_{Ic}$ are \emph{amplitude} parameters (which are different for different detectors)
and the vector $\bxi$ represents \emph{intrinsic} parameters of the signal (common for all detectors).
The log likelihood function for the signal \eqref{gensig} reads
\begin{align}
\label{logL2}
\ln\Lambda_I[x_I;A_{Is},A_{Ic},\bxi] &= \big(x_I(t)\big|s_I(t;A_{Is},A_{Ic},\bxi)\big)_I
\nonumber\\[1ex]&\quad
- \frac{1}{2}\big(s_I(t;A_{Is},A_{Ic},\bxi)\big|s_I(t;A_{Is},A_{Ic},\bxi)\big)_I,
\end{align}
where $x_I(t)$ represents the data collected by the $I$th detector
and where the scalar product $(\cdot|\cdot)_I$ is defined by
\be
(x|y)_I := 4\Re\int_0^\infty \frac{\tilde{x}(f)\tilde{y}^*(f)}{S_{nI}(f)}\mathrm{d}f.
\ee
Here $S_{nI}$ is the one-sided spectral density of the $I$th detector's noise
and tilde denotes Fourier transform.
Assuming that the noises in different detectors are independent of each other,
the log likelihood function $\ln\Lambda$ for the network of $N$ detectors reads
\be
\label{logLn}
\ln\Lambda[\mathbf{x};\mathbf{A}_{s},\mathbf{A}_{c},\bxi]
= \sum_{I=1}^N\ln\Lambda_I[x_I;A_{Is},A_{Ic},\bxi],
\ee
where $\mathbf{x}:=(x_1,\ldots,x_N)$, $\mathbf{A}_s:=(A_{1s},\ldots,A_{Ns})$,
and $\mathbf{A}_c:=(A_{1c},\ldots,A_{Nc})$.

To maximize the network log likelihood function \eqref{logLn} with respect to $A_{Is}$ and $A_{Ic}$ (for $I=1,\ldots,N$)
we find the unique solution to the equations
\be
\label{Amleqs2}
\frac{\partial \ln\Lambda}{\partial A_{Is}} = \frac{\partial \ln\Lambda_I}{\partial A_{Is}} = 0, \quad
\frac{\partial \ln\Lambda}{\partial A_{Ic}} = \frac{\partial \ln\Lambda_I}{\partial A_{Ic}} = 0.
\ee
The solution determines the maximum-likelihood estimators of the parameters $A_{Is}$ and $A_{Ic}$,
\be
\label{Amle2}
\hat{A}_{Is} = \frac{C_I(x_I|h_s)_I-M_I(x_I|h_c)_I}{D_I},
\quad
\hat{A}_{Ic} = \frac{S_I(x_I|h_c)_I-M_I(x_I|h_s)_I}{D_I},
\ee
where $C_I:=(h_c|h_c)_I$, $S_I:=(h_s|h_s)_I$, $M_I:=(h_s|h_c)_I$, and $D_I:=S_IC_I-M_I^2$.

After replacing in the log likelihood function \eqref{logL2}
the amplitudes $A_{Is}$ and $A_{Ic}$ by their estimators $\hat{A}_{Is}$ and $\hat{A}_{Ic}$,
we get the reduced likelihood function for the postmerger signal in the $I$th detector,
which we call the \emph{$\mathcal{P}_I$-statistic},
\begin{align}
\label{Fstat}
\mathcal{P}_I[x_I;\bxi] := \ln\Lambda_I[x_I;\hat{A}_{Is},\hat{A}_{Ic},\bxi]
= \frac{S_I(x_I|h_c)_I^2 + C_I(x_I|h_s)_I^2 - 2M_I(x_I|h_s)_I(x_I|h_c)_I}{2D_I}.
\end{align}
As the amplitudes $A_{I}$ for each detector are independent parameters,
the $\mathcal{P}$-statistic for the network is the sum of the $\mathcal{P}_I$-statistics for individual detectors,
\be
\label{Fstatn}
\mathcal{P}[\mathbf{x};\bxi] := \ln\Lambda[\mathbf{x};\hat{\mathbf{A}}_{s},\hat{\mathbf{A}}_{c},\bxi]
= \sum_{I=1}^N \mathcal{P}_I[x_I;\bxi].
\ee

The optimal SNR for the signal \eqref{gensig} seen in the $I$th detector is equal to
\be
\rho_I^2 = (s_I|s_I)_I = A_{Is}^2 S_I + 2 A_{Is}A_{Ic} M_I + A_{Ic}^2 C_I,
\ee
whereas the network SNR equals
\be
\label{SNRnet}
\rho_{\text{net}} = \sqrt{\sum_{I=1}^N \rho_I^2}.
\ee

In order to take into account coloured noise,
we perform whitening of the data by first dividing the Fourier transform of the data
by the square root of the spectral  density of noise
and then taking the inverse Fourier transform.
Thus the whitened data $x_{Iw}(t)$ in the $I$th detector is given by
\be
x_{Iw}(t) = \left(\mathcal{F}^{-1}\left\{\frac{\tilde{x}_I}{\sqrt{S_{nI}}}\right\}\right)(t),
\ee
where $\mathcal{F}^{-1}$ is the inverse Fourier transform.

We further assume that the filter functions $h_s$ and $h_c$ are narrowband:
from figure \ref{fig:spectra} one observes that the main spectral feature of the postmerger signal
is a few hundred Hz wide, whereas over the bandwidth of the data of a few kHz that we search
the spectral density of the noise is relatively flat (see figure \ref{fig:sen_all}).
Thus over the bandwidth of each filter the spectral density can be considered approximately
constant, $S_{nI}(f)\cong S_{Ic}=\text{const}$.
Consequently using Parseval's theorem
we approximate the scalar products $(\cdot|\cdot)_I$ present in Eq.\ \eqref{Fstat} as follows:
\begin{subequations}
\label{eq:sp_apr}
\begin{align}
(x_I|h_s)_I &\cong \frac{2\To}{\sqrt{S_{Ic}}}\av{x_{Iw}h_s},
\quad
(x_I|h_c)_I \cong \frac{2\To}{\sqrt{S_{Ic}}}\av{x_{Iw}h_c},
\\[1ex]
S_I &\cong \frac{2\To}{S_{Ic}}\av{h_s^2},
\quad
C_I \cong \frac{2\To}{S_{Ic}}\av{h_c^2},
\quad
M_I \cong \frac{2\To}{S_{Ic}}\av{h_s h_c},
\end{align}
\end{subequations}
where $\av{\cdot}$ denotes time averaging over the observational interval $\av{0;\To}$,
\be
\av{h} := \frac{1}{\To} \int_0^\To h(t)\,\mathrm{d}t.
\ee
With these approximations the $\mathcal{P}_I$-statistic has the form
\be
\label{eq:Pstatexact}
\mathcal{P}_I \cong \To \frac{S_0 F_{Ic}^2 + C_0 F_{Is}^2 - 2 M_0 F_{Ic} F_{Is}}{D_0},
\ee
where
\begin{subequations}
\begin{align}
F_{Is} &:= \av{x_{Iw}h_s},
\quad
F_{Ic} := \av{x_{Iw}h_c}
\\[1ex]
C_0 &:= \av{h_c^2},
\
S_0 := \av{h_s^2},
\
M_0 := \av{h_ch_s},
\
D_0 := S_0 C_0 - M_0^2.
\end{align}
\end{subequations}

Let us now assume that the filter functions $h_c$ and $h_s$ have the following form
\begin{subequations}
\label{hchs}
\begin{align}
h_c(t;\bxi) &:= a(t;\bxi)\cos\Phi(t;\bxi),
\\[1ex]
h_s(t;\bxi) &:= a(t;\bxi)\sin\Phi(t;\bxi),
\end{align}
\end{subequations}
where the phase $\Phi(t;\bxi)$ has very many oscillations over the observational interval
whereas $a(t;\bxi)$ is a slowly varying
(compared to the typical period of oscillation of the phase $\Phi$) function of time.
Then with good accuracy, we have
\be
C_0 \cong S_0 \cong \frac{1}{2}\av{a^2},
\quad
M_0 \cong 0.
\ee
With these approximations the $\mathcal{P}_I$-statistic \eqref{eq:Pstatexact}
can be expressed as
\be
\label{eq:Pstat}
\mathcal{P}_I \cong \frac{2}{\To} \frac{|\tilde{F_I}|^2}{\av{a^2}},
\ee
where
\be
\label{FT1}
\tilde{F}_I := \int_0^\To x_{Iw}(t) a(t;\bxi)\exp(-i \Phi(t;\bxi))\,\md t.
\ee

For the model of the GW signal defined by Eqs.\ \eqref{RF7}--\eqref{RF8}
with the phase $\Phi$ given by Eq.\ \eqref{Phi},
the $\mathcal{P}_I$-statistic can be computed from Eq.\ \eqref{eq:Pstat},
in which
\be
\av{a^2} = \frac{1}{\To}\int_0^\To e^{-2ot}\,\mathrm{d}t
= \frac{1}{2o\To}\big(1-e^{-2o\To}\big),
\ee
and
\be
\label{FT}
\tilde{F}_I = \int_0^\To x_{Iw}(t) \exp(-t/\tau) \exp(-2\pi i f t - 2\pi i \gamma t^2)\,\md t.
\ee
We see that $\tilde{F}_I$ is the Fourier transform of the function
\be
F_I(t) := x_{Iw}(t)\exp(-t/\tau) \exp(- 2\pi i \gamma t^2),
\ee
which is the amplitude and frequency drift demodulated data.
For discrete in-time data, the integral \eqref{FT} becomes the discrete Fourier transform of the function $F_I$,
which can be computed using the FFT algorithm.

\section{Monte Carlo simulations: LIGO data}
\label{sec:injections}

To test how effective the matched-filtering statistic derived in the previous section is,
we have performed Monte Carlo simulations by injecting postmerger waveforms
from the CoRe numerical database to the LIGO data.
We have generated the responses of the LIGO detectors to postmerger waveforms
as described in section \ref{sec:response} with orientation fixed to GW170817 signal.
In our simulations, we have used public LIGO data from the Gravitational Wave Open Science Center
(GWOSC)\footnote{{\tt https://gwosc.org}.} \cite{RICHABBOTT2021100658},
containing the BNS merger event GW170817.
We have not used Virgo data because their amplitude spectral density was almost an order of magnitude greater
than that of the LIGO detectors for frequencies $\geq 1$~kHz
and they would contribute very little to the network signal-to-noise ratio of the signal.

We have done our injections off-source in time, assuming the sky position of the GW170817 merger.
We have taken the 14-second-long stretch of data starting 2 seconds
after the time $t_m$ of the GW170817 merger ($t_m = 1187008882.430\pm0.002$~s \cite{Abbott_2017b}).
For each injection, we randomly selected a starting time within the 14-second-long stretch.
We have performed the search for the network of Hanford (H1) and  Livingston (L1) detectors.
We have performed our simulations separately for each of the 24 waveforms
consistent with the GW170817 event \cite{Nedora_2021}.
We have done simulations for a range of optimal network SNRs [see Eq.\ \eqref{SNRnet}]
starting from 0 (no signal added) to 25 with the step of 1.
We have scaled the amplitudes of the injected signals to obtain the desired optimal SNR.
For each SNR we have made 1000 signal injections to the LIGO data. 
In our simulations, the length of the data we analyzed was equal to the length of the postmerger waveform.
To detect the signals and estimate their parameters we have used
the ${\mathcal P}_I$-statistic given by Eq.\ \eqref{eq:Pstat}
with filters $h_c$ and $h_s$ given by Eqs.\ \eqref{RF8}.
In the search we have chosen the following grid in the 3-dimensional parameter space $(o,f,\gamma)$:
for $(f,\gamma)$ subspace we have used the optimized grid constructed
in \cite{pisarski:2011} with minimal match parameter $m = 0.9^{1/3}$,
and for the parameter $\tau$ we have used a uniform grid in its inverse $o = 1/\tau$
with spacing $\Delta o=0.01$.
We have searched the frequency range $f \in\langle1150;4000\rangle$~\text{Hz}. For the remaining two
parameters $\gamma$ and $\tau$ we have selected ranges depending on the injected waveform
in the following way. For each case, we have added a waveform with the high SNR of 100 to the data.
We have estimated the parameters $\gamma$ and $\tau$.
Then in the simulation, we searched for the signals $\pm 3$ grid points
around the estimated parameters of $\gamma$ and $\tau$.
This was sufficient to estimate the parameters of the injected signals.
In the analysis of the LIGO data presented in the next section,
we search the full ranges of the parameters $\gamma$ and $\tau$.

\begin{figure}
\includegraphics[width=\textwidth]{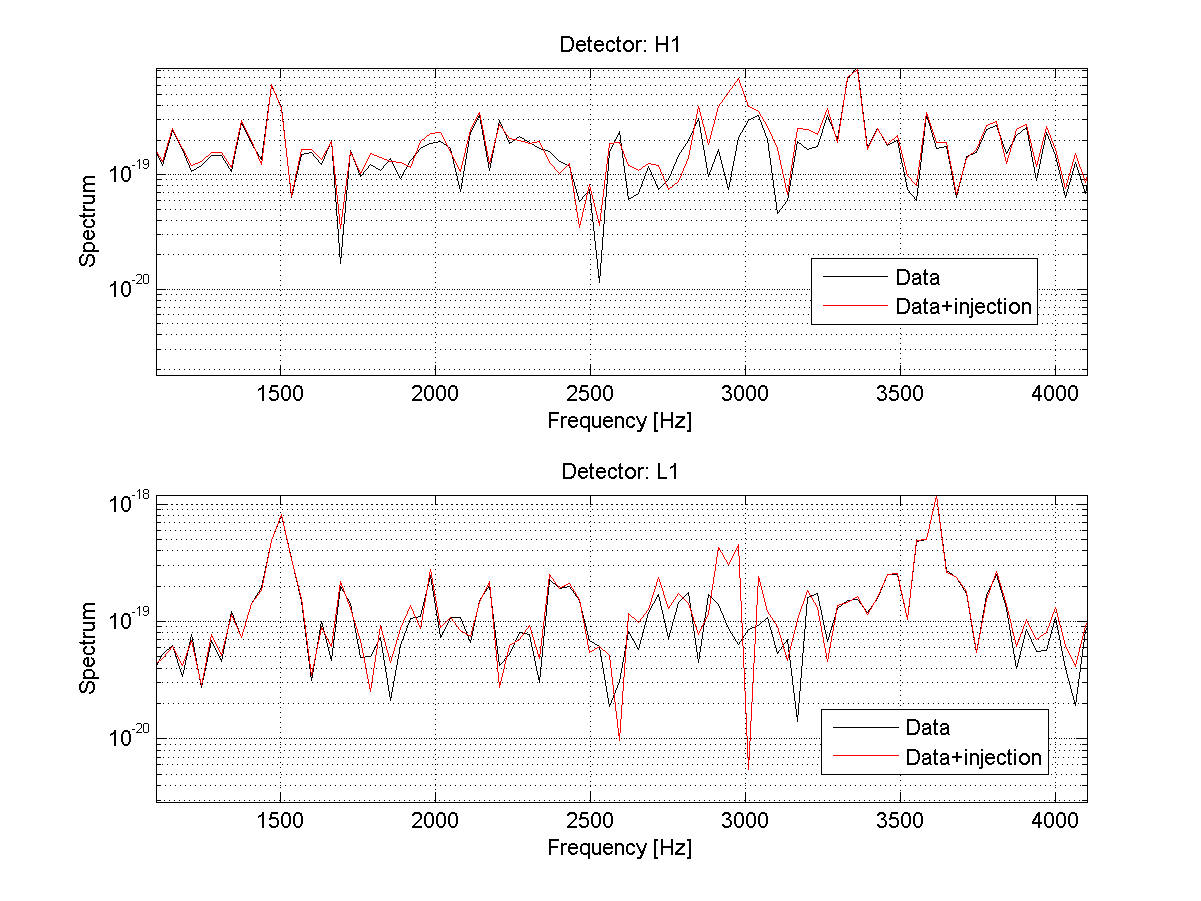}
\caption{Spectrum of the LIGO data compared to the spectrum of the same data with postmerger
waveform from the numerical simulation THC0105 added. The amplitude of the waveform is scaled 
so that the signal has network SNR $\rho_{\text{net}}=10$.
The top panel is for the LIGO Hanford detector
whereas the bottom panel is for the Livingston detector.}
\label{fig:SigAdded}
\end{figure}

\begin{figure}
\includegraphics[width=\textwidth]{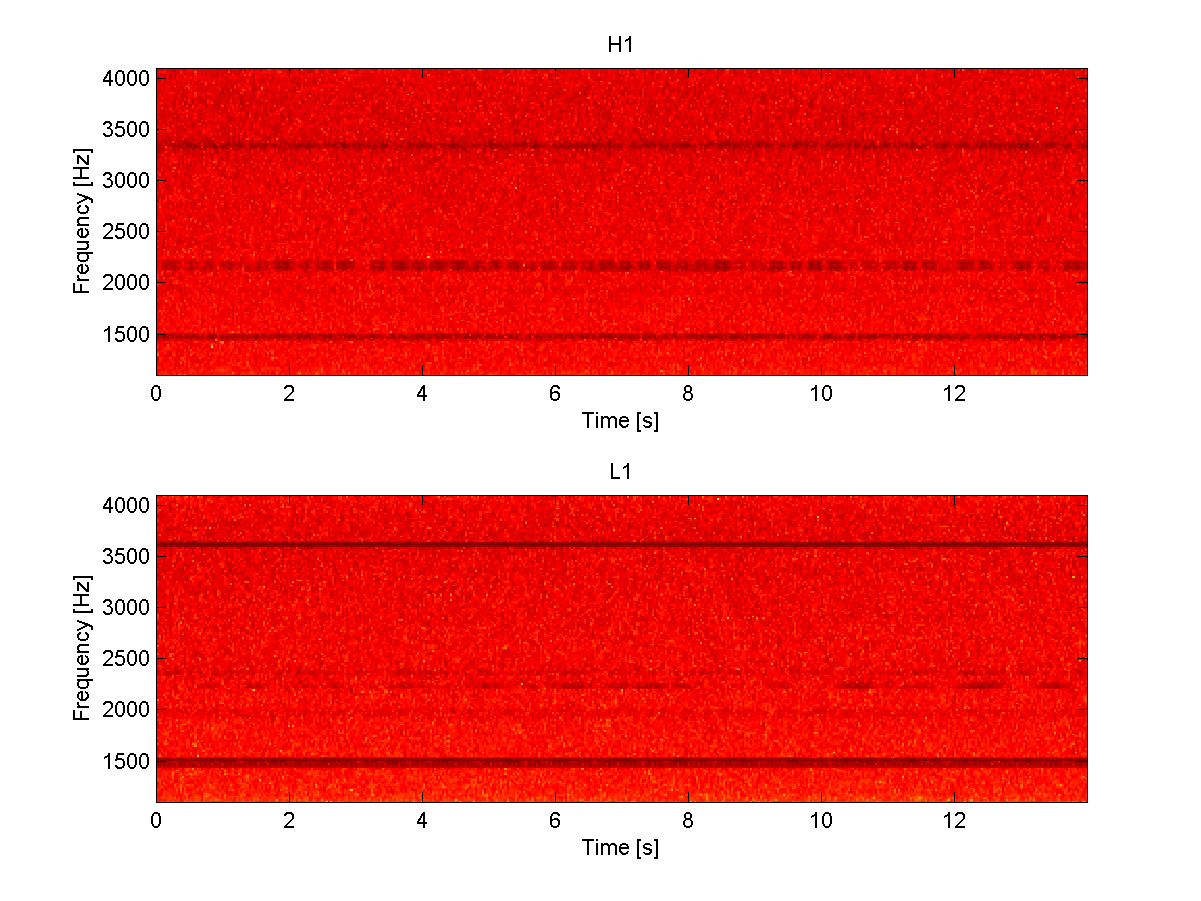}
\caption{Spectrograms of the 14-second-long stretches of LIGO data taken 2 seconds after the GW170817 merger.
Top panel---Hanford detector data. Bottom panel---Livingston detector data.}
\label{fig:specgramH1}
\end{figure}

As the noise in a detector is not white we perform whitening of the data  using an estimate
of the noise spectral density.  We need to consider the following effect.
For signals with SNR in our investigated SNR range
the presence of the signal affects the spectral density of the noise.
This is illustrated in figure \ref{fig:SigAdded},
where we compare the spectra of the two LIGO detectors data
with data containing GW signal of the network SNR equal to 10.
To take this effect into account we divide the 14-second-long stretch of LIGO detectors data
into overlapping segments and we estimate the spectral density of each segment.
Then we take the average of the spectra and these averaged spectra are used
to whiten the noise in each detector before applying our statistic.
In figure \ref{fig:specgramH1} we present the spectrograms of the 14-second-long stretches
of data of the two LIGO detectors.

Our simulations proceed as follows.
After adding a signal from the CoRe database to LIGO detectors data with a given network SNR,
we evaluate the approximate ${\mathcal P}$-statistic [from Eq.\ \eqref{eq:Pstat}]
over the grid described above.
We find the maximum and record the parameters of the maximum.
Then we perform the second step to find more accurately the maximum of the ${\mathcal P}$-statistic
over our 3-dimensional parameter space. We apply the Nelder-Mead maximization algorithm \cite{Nelder_1998}
and we maximize the exact ${\mathcal P}$-statistic given by Eq.\ (\ref{eq:Pstatexact}).
We take as the initial values for the maximization procedure
the values of the parameters from the maximum of the grid search.
The values of the parameters $f$, $\tau$, and $\gamma$
for which ${\mathcal P}$ is maximum
are maximum likelihood estimators that we denote by $\hat{f}$, $\hat{\tau}$, and $\hat{\gamma}$,
respectively.
The estimators $\hat{f}$, $\hat{\tau}$, and $\hat{\gamma}$
are then used in Eqs.\ (\ref{Amle2}) to obtain the maximum likelihood estimators
of the amplitudes $A_c$ and $A_s$.
Finally, we reconstruct the postmerger signal using the formula
\be
\label{eq:sigrec}
\hat{s}(t) = \hat{A}_c\exp(-t/\hat{\tau}) \cos(2\pi \hat{f} t + 2\pi \hat{\gamma} t^2)
+ \hat{A}_s\exp(-t/\hat{\tau}) \sin(2\pi \hat{f} t + 2\pi \hat{\gamma} t^2).
\ee

\begin{figure}
\includegraphics[width=\textwidth]{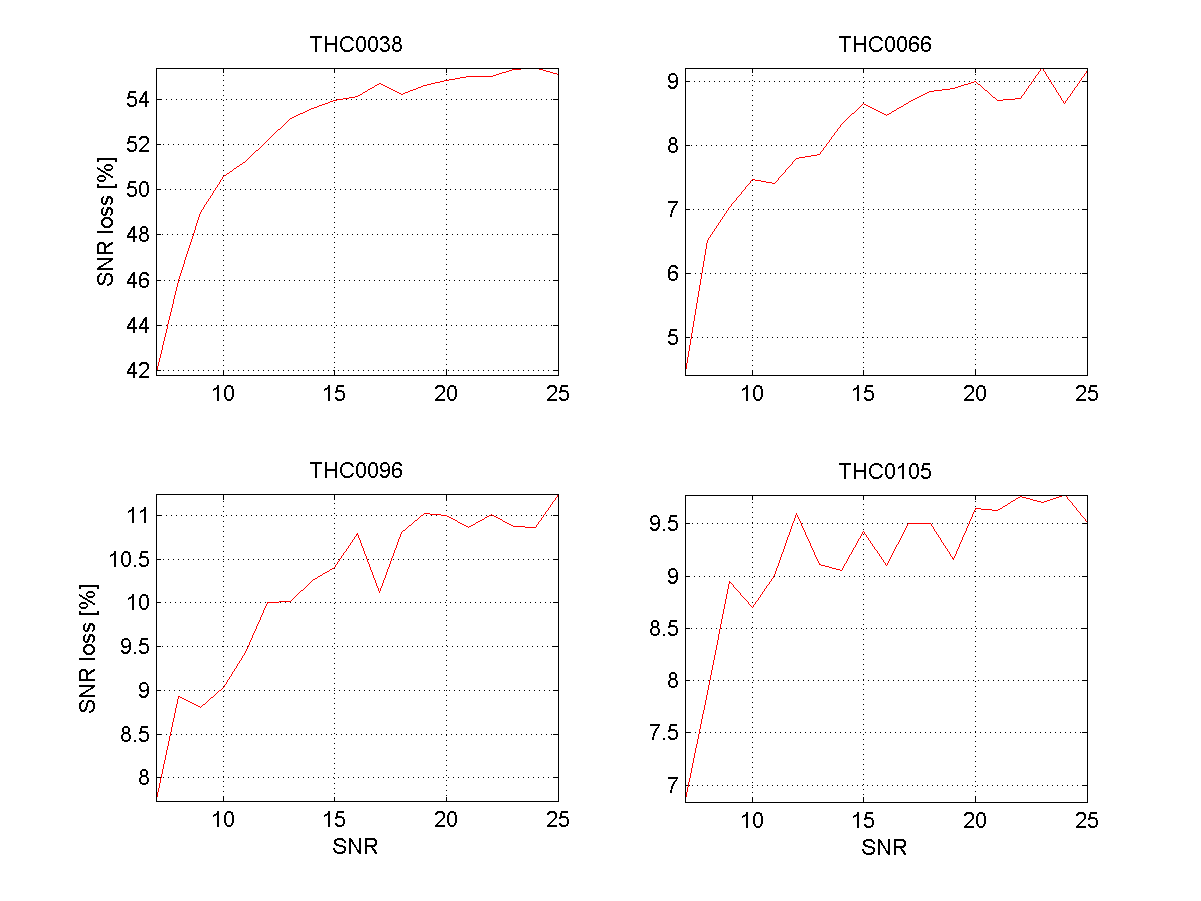}
\caption{Average fractional loss of the SNR as a function of the injected SNR
for four different numerical simulations of the postmerger signal.}
\label{fig:snrlossGW17}
\end{figure}

In figure \ref{fig:snrlossGW17} we have presented the quality of signal detection 
by plotting the SNR loss as a function of the SNR of the injected signal.
The fractional SNR loss $l$ is defined as
\be
l  =  \frac{\rho_{\text{net}} - \rho_r}{\rho_{\text{net}}},
\ee
where $\rho_{\text{net}}$ is the optimal network SNR [given by Eq.\ \eqref{SNRnet}] of the injected signal
and $\rho_r$ is the recovered SNR.
For each $\rho_{\text{net}}$ we calculate $\rho_r$ as the mean of the recovered SNRs of the 1000 injected signals. 

\begin{figure}
\includegraphics[width=\textwidth]{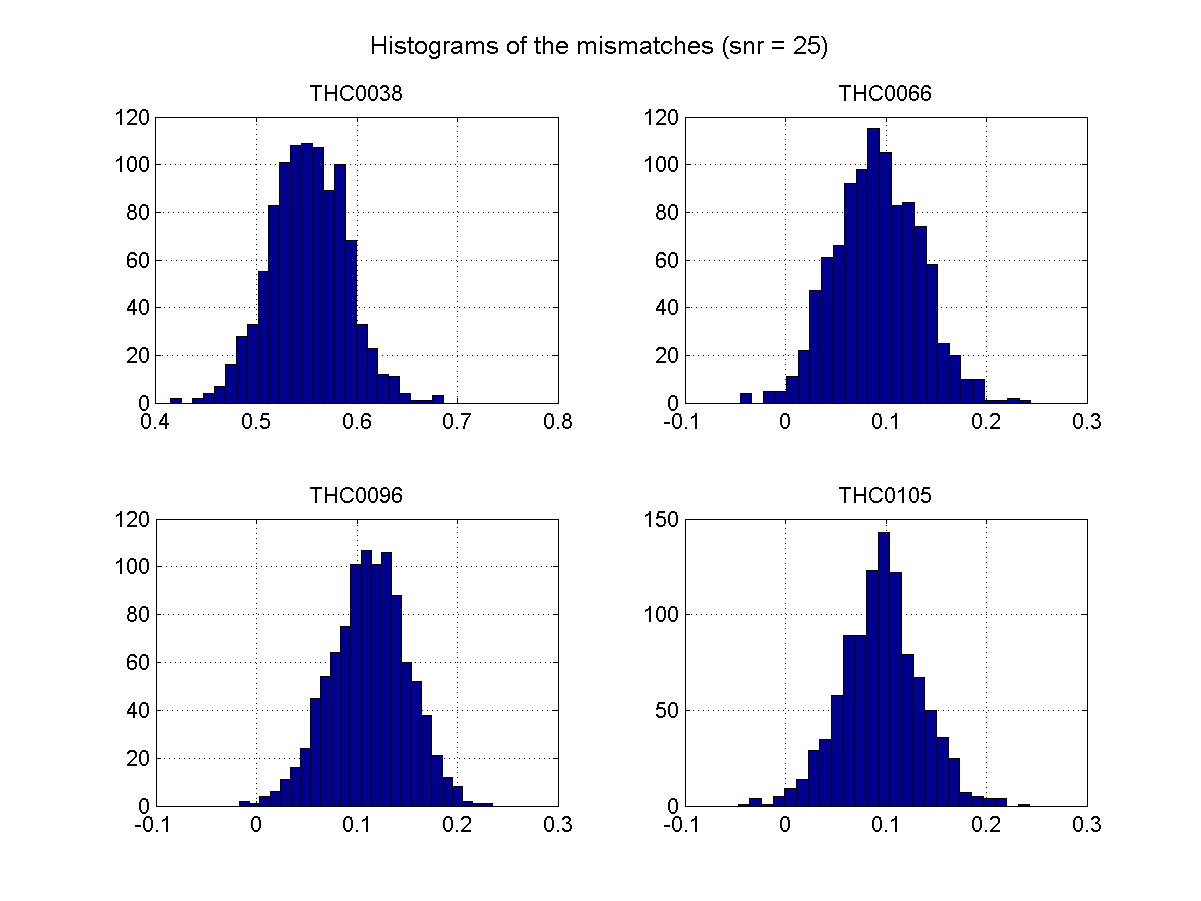}
\caption{Histograms of the SNR mismatches for the four simulations.}
\label{fig:hist_25}
\end{figure}

\begin{figure}
\includegraphics[width=\textwidth]{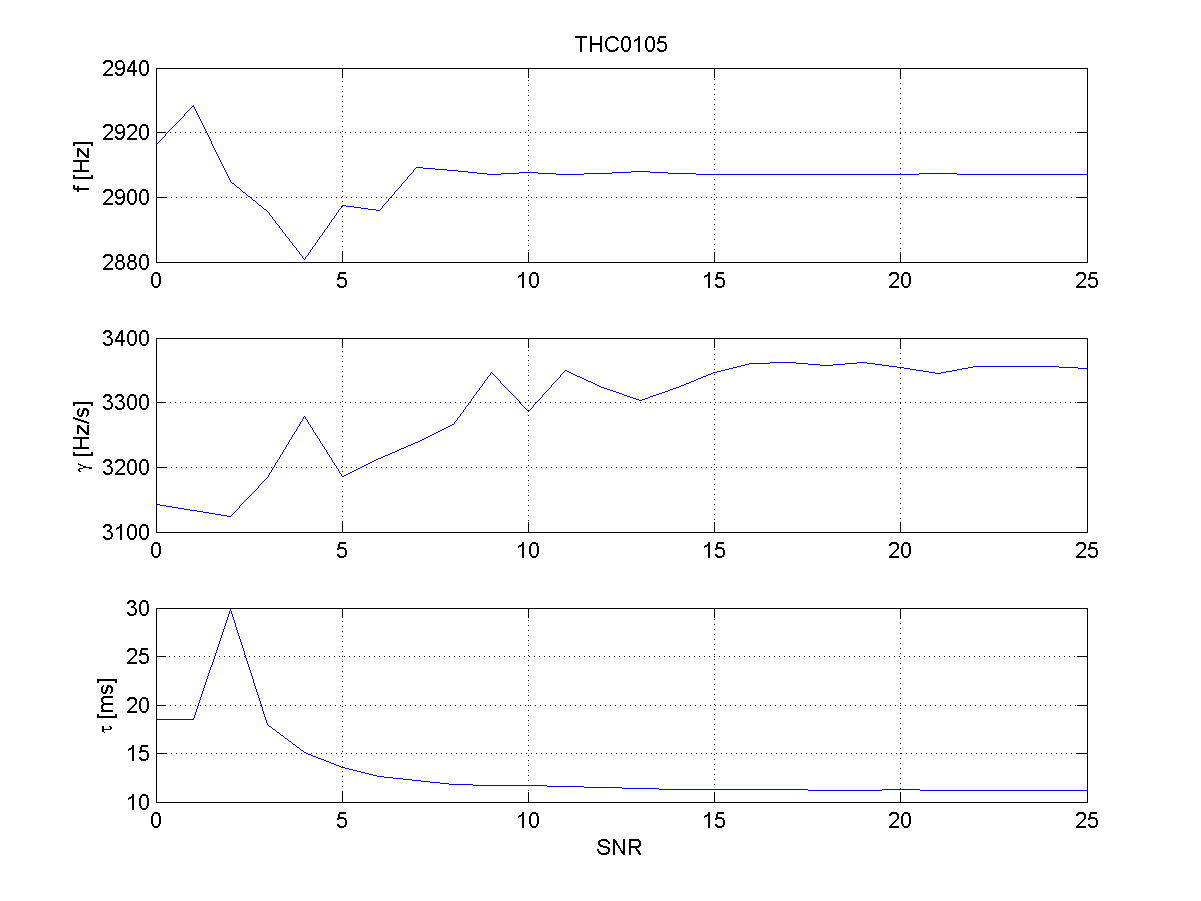}
\caption{Estimators of the three intrinsic parameters $f$, $\gamma$, and $\tau$
as functions of the SNR of the injected signal for the case of the THC0105 waveform.}
\label{fig:parestTHC0105}
\end{figure}

In figure \ref{fig:hist_25} we have plotted histograms of the mismatches defined 
as the relative difference between the injected SNR and the recovered one.
The mismatches are plotted for the highest $\text{SNR}=25$ for which we inject signals.
In figure \ref{fig:parestTHC0105} we have presented the values of the estimators
of the three intrinsic parameters $f$, $\gamma$, and $\tau$ of the signal
as a function of the SNR of the injected waveform
for the case of the numerical model THC0105. For each SNR the estimators are calculated
as means of the estimates of the parameters from 1000 injections.

\begin{figure}
\includegraphics[width=\textwidth]{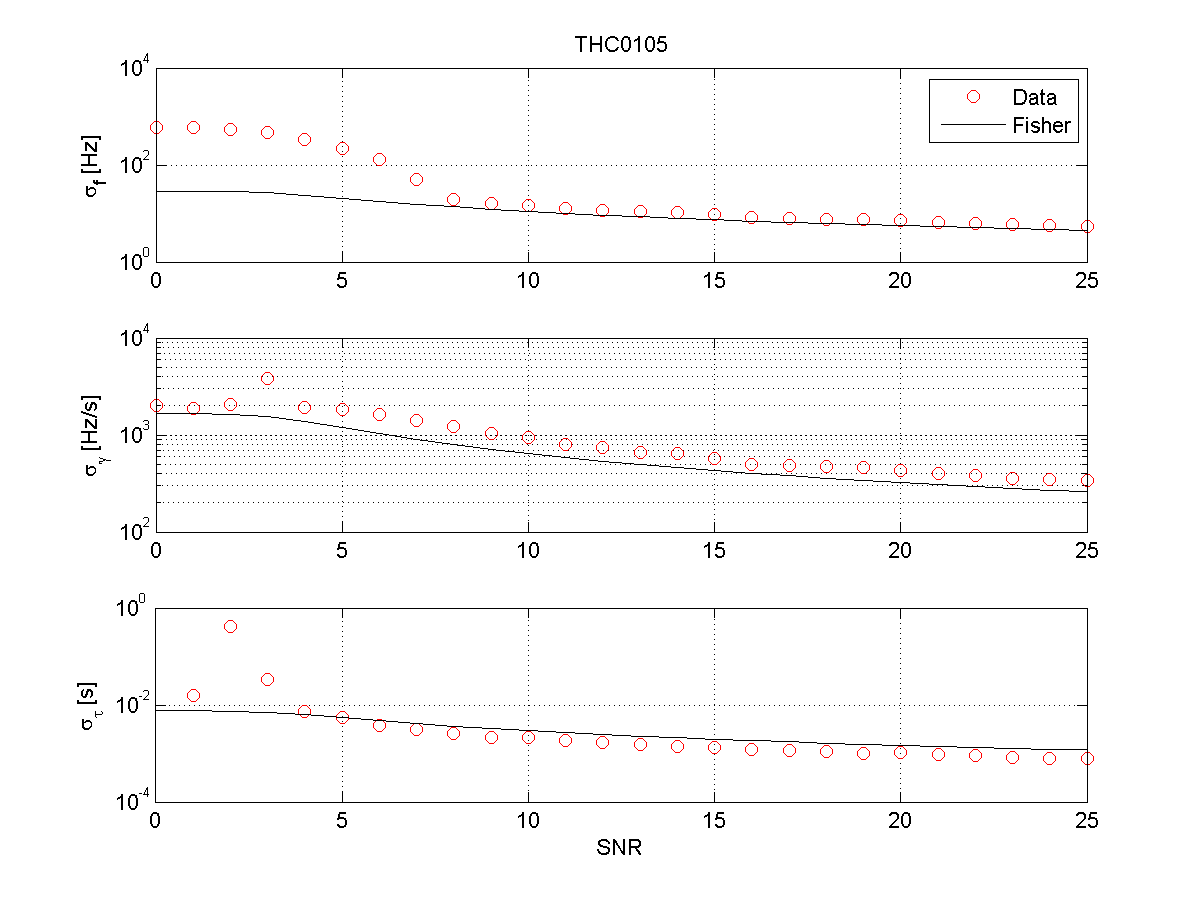}
\caption{Standard deviations of the estimators of the intrinsic parameters $f$, $\gamma$, and $\tau$
as functions of the SNR of the injected signal for the case of the THC0105 waveform.
The estimates from the injections are marked by circles. Continuous lines
are theoretical estimates of the standard deviations computed from the Fisher matrix.}
\label{fig:sigestTHC0105}
\end{figure}

In figure \ref{fig:sigestTHC0105} we have presented the errors of the estimators of the parameters
as a function of the SNR of the injected signal.
For each SNR the errors are calculated as standard deviations of estimators of the parameters from 1000 injections.
We compare these errors with errors
estimated from the Fisher matrix (see \ref{app:FAP} for details).
In the calculation of the Fisher-matrix errors,
we use the value of the recovered SNR $\rho_r$ of the injections.

In table \ref{tab:simresGW17} we present the results of the simulations
when the injected signal has the optimal network SNR $\rho_{\text{net}}$ equal to 25.
We show the SNR loss, estimated frequency $f$, frequency drift $\gamma$, and damping time $\tau$,
together with their errors.
The estimated value of any parameter is the mean of the estimated parameters taken from the 1000 simulations
and the error is the standard deviation of the 1000 estimated parameters.
We also show the maximum frequencies $f_{\text{peak}}$ of the spectra of the waveforms.

\begin{table}
\caption{\label{tab:simresGW17}
The results of simulations performed for injected signals
to the LIGO data with the optimal SNR equal to 25.}
\lineup
\begin{indented}
\item[]\begin{tabular}{ccccccccc}
\br
PM
& SNR
& $f_{\rm peak}$
& $f_{\rm est}$
& $\sigma_{f}$
& $\gamma$
& $\sigma_{\gamma}$
& $\tau$
& $\sigma_{\tau}$
\\[0ex]
waveform
& loss [\%]
& [Hz]
& [Hz]
& [Hz]
& [Hz/s]
& [Hz/s]
& [ms]
& [ms]
\\ \mr
THC0038 & 55.1 & 3076 &  2846.7  &  13.3 & \09459.9  & \0709.7  &  14.2  &   1.8\\
THC0041 & 31.2 & 2448 &  2660.1  &  14.0 & \-22007.3 & 1964.2  &   \05.6  &   0.4 \\
THC0042 & 30.5 & 2460 &  2608.1  &  14.4 & \-14587.1 & 1787.1  &   \06.3  &   0.5 \\
THC0043 & 24.6 & 2468 &  2563.5  &  11.4 & \-14053.6 & 1340.5  &  \07.1  &   0.5\\
THC0052 & 22.6 & 2992 &  3061.3  &  20.6 & \0\-7351.0 & 2199.6  &   \06.3  &   0.8 \\
THC0053 & 27.2 & 3104 &  3078.0  &   \05.3 & \01134.5 &  \0250.4  &  10.7  &   1.1 \\
THC0054 & 33.2 & 3088 &  2937.9  &  14.9 & 10097.0 & 1347.2  &   \09.3 &    1.1 \\
THC0055 & 25.7 & 2448 &  2509.5  &  13.3 & \-13536.9 & 1706.8  &   \06.6  &   0.5 \\
THC0063 & 12.8 & 3052 &  2995.0  &   \04.7 & \03146.7 &  \0206.0 &   13.7  &   1.0 \\
THC0064 & 26.1 & 2932 &  2983.6  &  11.0 & \0\-5423.7 & 1155.5  &   \08.0  &   0.5 \\
THC0065 & 50.2 & 3108 &  2819.4  &   \09.8 & 15151.4 &  \0793.4  &  12.6  &   1.2  \\
THC0066 & \09.2  & 2688 &  2692.3 &   30.7 & \0\-1459.8 & 6368.1  &   \05.0  &   0.4\\
THC0077 & 28.3 & 3012 & 2955.8  &   5.0 & \03735.1 &  \0214.5  &  16.9  &   1.4 \\
THC0079 & 19.2 & 2976 &  2853.0  &  18.4 & 16479.9 & 1880.7  &   \09.7  &   0.8\\
THC0080 & 26.6 & 2960 &  2807.6  &  26.5 & 14346.3 & 1987.1  &  18.7  &   3.6\\
THC0081 & 22.5 & 2912 &  2758.8 &   13.8 & 14755.1 &  \0894.7 &   12.0 &    0.7\\
THC0082 & 17.3 & 3024 &  2907.4  &  11.0 & 11364.7 &  \0961.7  &   \08.9  &   0.6\\
THC0083 & 17.1 & 3168 &  3025.7  &  25.6 & 15054.1 &  2931.1 &   10.3  &   1.0\\
THC0087 & 32.9 & 3068 &  3007.0  &   \07.1 & \03697.8 &  \0246.6  &  23.5  &   1.7\\
THC0089 & 34.4 & 3328 &  3228.2  &  19.1 & \09524.1 & 1831.3  &  10.7  &   1.4\\
THC0091 & 33.1 & 3360 &  3271.6  &  40.2 & \08663.4 & 3139.0  &   \05.7  &   0.6\\
THC0095 & 34.0 & 3456 &  3367.5  &   31.7 & \09226.5 & 2511.7  &   \07.3  &   0.7\\
THC0096 & 11.2  & 3000 &  2929.3  &   \06.7 & \04589.1  & \0316.9 &   17.0  &   1.1\\
THC0105 & \09.5 & 2968 &  2907.1  &   \05.5 & \03352.3  & \0335.8  &  11.1  &   0.8\\
\br
\end{tabular}
\end{indented}
\end{table}

\begin{figure}
\includegraphics[width=\textwidth]{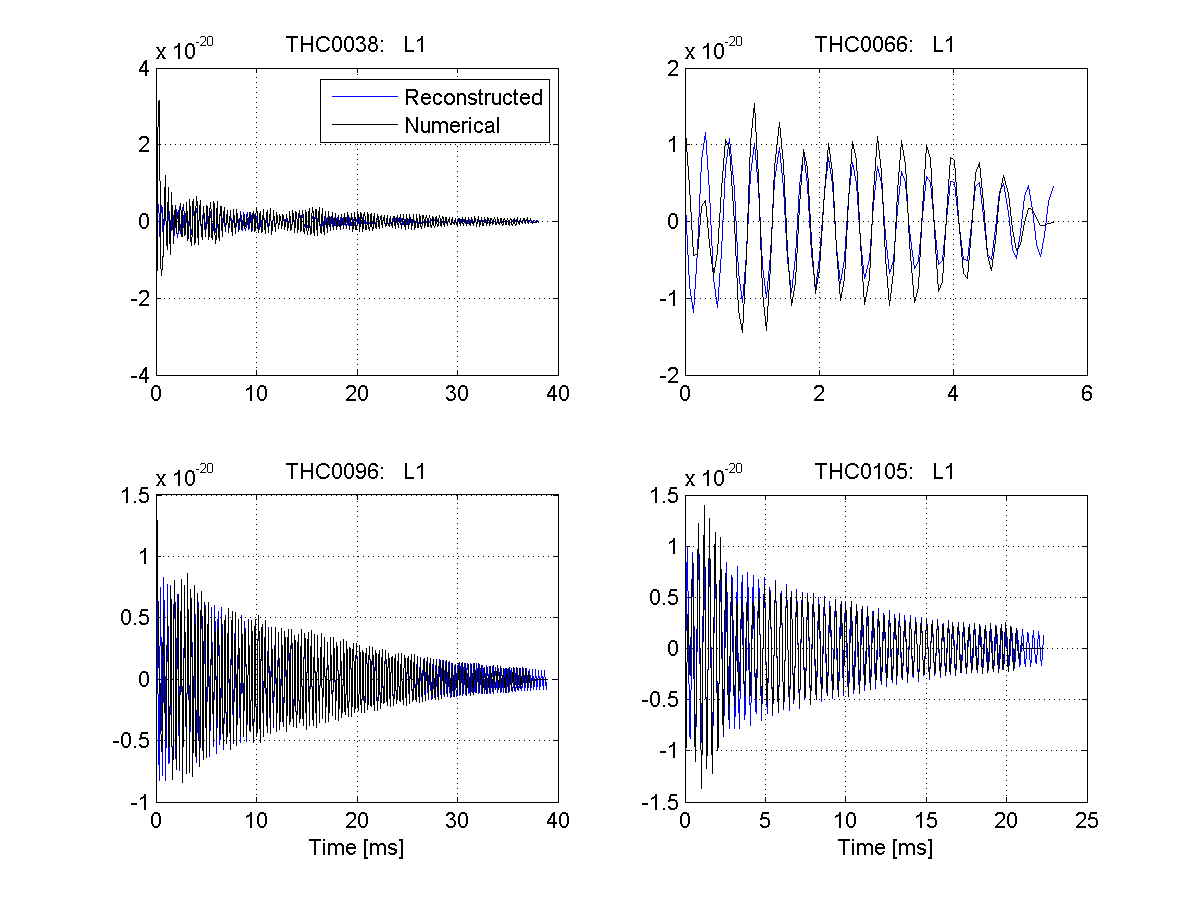}
\caption{Comparison of the responses to the injected waveforms for $\text{SNR}=25$
with the reconstructed responses calculated from Eq.\ \eqref{eq:sigrec}
with parameters estimated from the simulations.}
\label{fig:rec_data_25}
\end{figure}

\begin{figure}
\includegraphics[width=\textwidth]{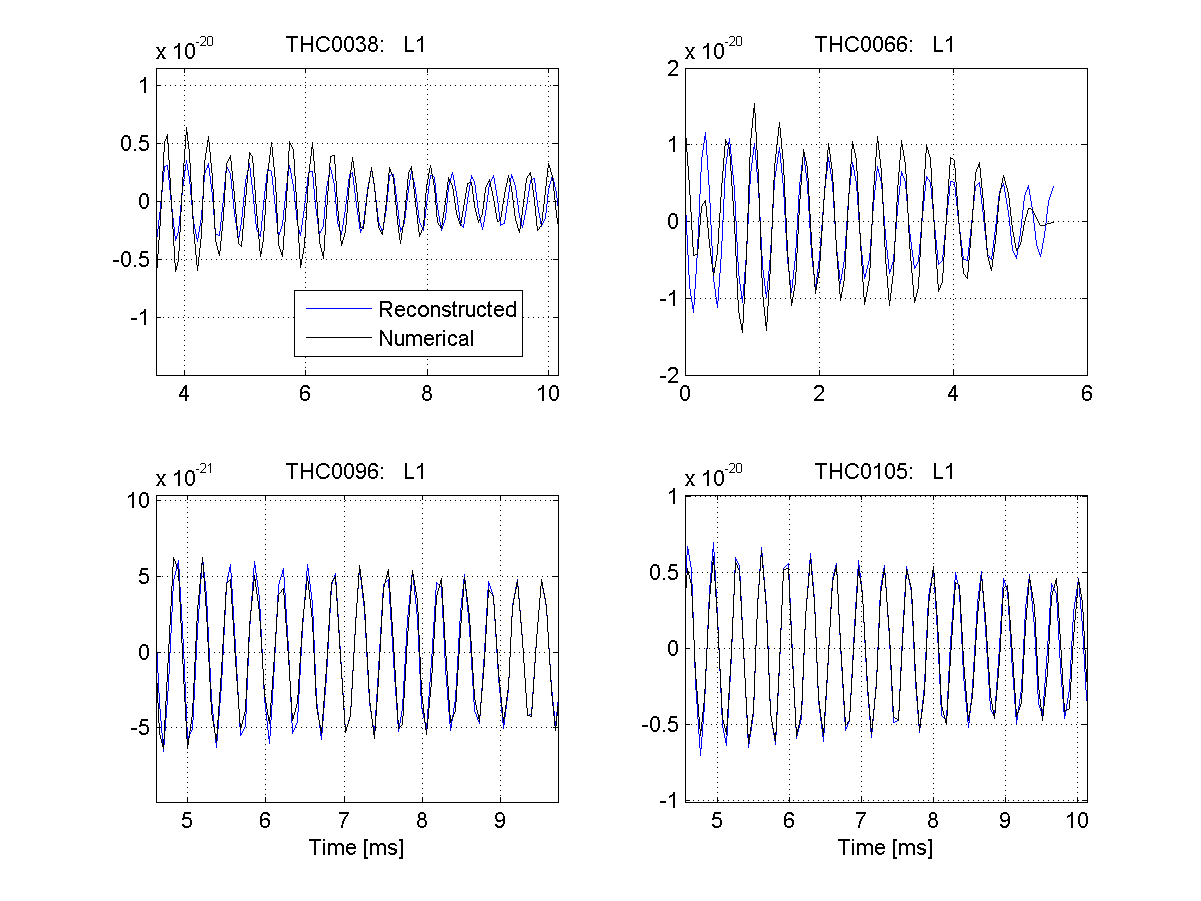}
\caption{The same as in figure \ref{fig:rec_data_25}
but with the time interval restricted to $\langle5;10\rangle$~ms.}
\label{fig:rec_data_25_z}
\end{figure}

In figure \ref{fig:rec_data_25} we have compared the injected responses of the Hanford detector 
to the waveforms with the reconstructed responses obtained from Eq.\ \eqref{eq:sigrec}.
Figure \ref{fig:rec_data_25_z} is a zoom of figure \ref{fig:rec_data_25}
to  better illustrate the accuracy of our detector response reconstruction.

The postmerger signals after the BNS merger exhibit a very complex morphology however
our templates based on the single-damped sinusoid model provide a reasonable detectability of these signals 
with matched filtering technique. From Table \ref{tab:simresGW17},
we see that for half of the 24 waveform models considered the SNR loss with respect 
to a perfectly matched filter is around 30\%, 4 cases exhibit SNR loss 
of around 10\%, and 6 of around 20\%. For only two cases the SNR loss is around 50\%. 
The estimated frequency $f_{\rm est}$ of the postmerger
signal agrees very well with the maximum frequency $f_{max}$ of the spectrum of the signal.
Also, for SNR = 25 (the table has been done for this SNR value)
$f_{\rm est}$ is very well estimated with 1-$\sigma$ error of less than 1\%.
The damping time $\tau$, giving the characteristic length of the postmerger signal,
varies from around 5~ms to around 25~ms and is not correlated with the SNR loss of our model.
The damping time is estimated with around 10\% accuracy for SNR = 25. The frequency drift 
$\gamma$ shows a very wide variation and large errors. The estimated values of $\gamma$ 
are probably affected by other components of the postmerger signal that interfere with the main 
component modelled by our template. The smaller SNR loss tends to give smaller values
of the frequency drift.

\section{Search of the LIGO data for the postmerger signal of the GW170817 event}
\label{sec:searchGW17}

\begin{figure}
\includegraphics[width=\textwidth]{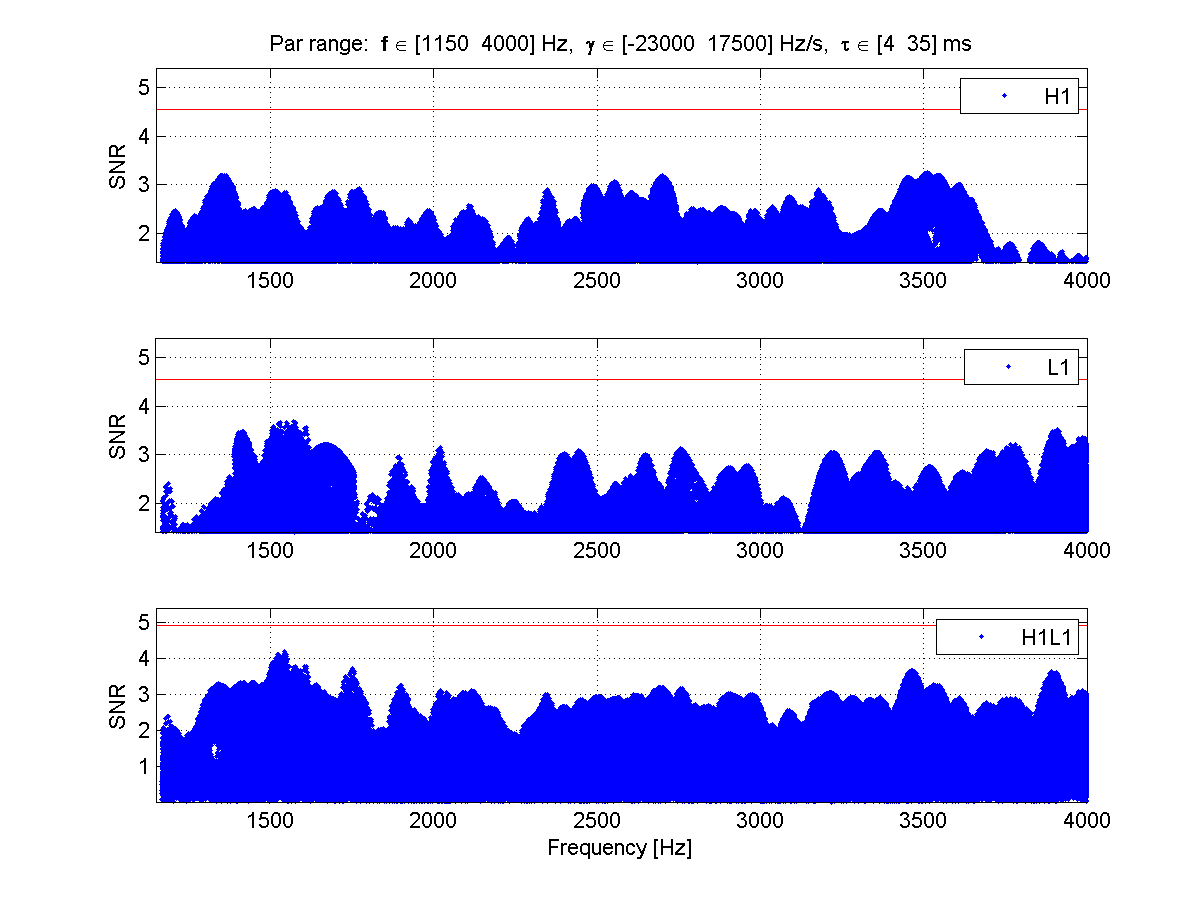}
\caption{Triggers from the postmerger signal search after the GW170817 merger.
The top and middle panels are triggers from Hanford and Livingston detector searches.
The bottom panel is for the search of the network of the two LIGO detectors.
The red horizontal lines denote thresholds corresponding to 1\% false alarm probability.}
\label{fig:cand_GW17}
\end{figure}

We have performed a search for the postmerger signal after the GW170817 event
in the network of the two LIGO detectors using the statistic presented in section \ref{sec:det} in the same way,
as we searched for the postmerger waveforms from the CoRe database described in the previous section.
In particular to estimate spectral density we have used the 14-second-long stretch of data taken 2~s after the merger.
This ensures that the estimate of the spectral density is in no way contaminated by any postmerger signal present. 
For the analysis, we have taken 35~ms of data in each LIGO detector after the time of the merger.
We have performed the search in the Hanford and Livingston data separately and in the network of the two detectors.
We have searched the following ranges of the parameters $\tau$, $f$, and $\gamma$ of our templates:
\begin{subequations}
\begin{align}
\tau &\in \langle4;35\rangle~\text{ms},
\\[1ex]
f &\in \langle1150;4000\rangle~\text{Hz},
\\[1ex]
\gamma &\in \langle-23000;17500\rangle~\text{Hz/s}.
\end{align}
\end{subequations}
The ranges of the parameters were guided by the parameters of the postmerger waveforms
obtained from the analysis in the previous section.
The estimates of parameters $\tau$, $f$, and $\gamma$ reported in Table \ref{tab:simresGW17}
are well within the search ranges given above.
For the search in the individual detectors, we set the threshold on the $\mathcal{P}$-statistic
equal to 1 and for the network search the threshold equal to 2.
The triggers obtained in our searches are presented in figure \ref{fig:cand_GW17}. 

We see that none of the triggers crossed the threshold corresponding to 1\% false alarm probability.
Thus no significant postmerger signal has been detected. This was expected for the event GW170817 
and the result is consistent with the analysis performed in \cite{Abbott_2017} 
which used the coherent Wave Burst (cWB) pipeline \cite{Klimenko-2008,PhysRevD.93.042004}.
The dominant part of the postmerger signal is a periodic signal with frequency $f$.
As our statistic involves search for this periodicity using the Fourier transform
we expect that small transients present in the signal and an approximate estimate of its starting time
will not degrade significantly the signal-to-noise ratio achieved with our method.

\section{Sensitivity of the search}
\label{sec:senGW17}

We express the sensitivity of our search to a given waveform model by the quantity $h_{\rm rss}^{50\%}$,
which is the root-sum-squared strain amplitude of signals
which are detected with 50\% efficiency \cite{Sutton2013,Abbott_2017}.
To calculate $h_{\rm rss}^{50\%}$ we set a detection threshold
corresponding to a false-alarm probability of $10^{-4}$.
In \ref{app:FAP} we have presented how false alarm probability and the corresponding threshold are calculated.
The quantity $h_{\mathrm{rss}}$ is defined as
\begin{equation}
\label{eq:hrss}
h_{\mathrm{rss}} = \sqrt{2\int_{f_{\textrm{min}}}^{f_{\textrm{max}}} \left( |\tilde{h}_+(f)|^2 + |\tilde{h}_{\times}(f)|^2 \right) \mathrm{d}f},
\end{equation}
where $f_{\mathrm{min}}$ and $f_{\mathrm{max}}$ are respectively
the minimum and maximum frequencies over which the search is performed.

\begin{figure}
\includegraphics[width=\textwidth]{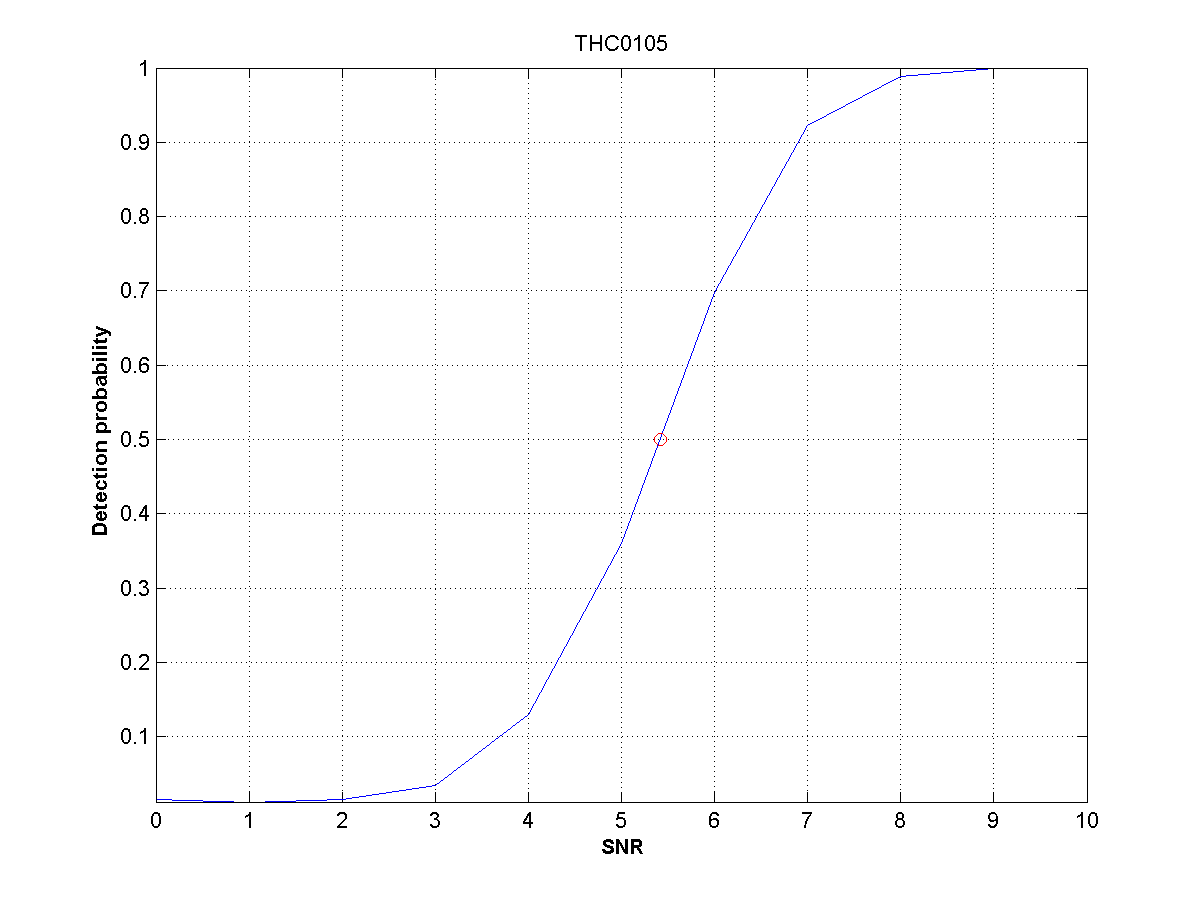}
\caption{Probability of detection of the waveform THC0105
in the network of LIGO detectors in the data following the GW170817 event
as a function of the injected SNR for the threshold corresponding 
to $10^{-4}$ false alarm probability.
By round circle we denote interpolation of SNR
to 50\% detection probability. In this case $\rho^{50\%} \cong 5.4$.}
\label{fig:drss_THC0105}
\end{figure}

To calculate the sensitivity $h_{\rm rss}^{50\%}$ for a given waveform,
we first obtain the SNR $\rho^{50\%}$ for the waveform
at which we get 50\% probability of detection with the threshold corresponding
to the false-alarm probability equal to $10^{-4}$.
We obtain $\rho^{50\%}$ by estimating the probability of detection
from our Monte Carlo simulations presented in section \ref{sec:injections} 
for each SNR at which we inject the waveform,
and then by interpolating we obtain the SNR $\rho^{50\%}$ corresponding to the 50\% probability of detection.
This is illustrated in figure \ref{fig:drss_THC0105} for the case of the waveform THC0105
where $\rho^{50\%}\cong5.4$.

We then scale the amplitudes of the two polarizations $h_+$ and $h_{\times}$ by the amplitude
which is the ratio of $\rho^{50\%}$ and the network SNR for the waveform given 
by the sixth column of table \ref{tab:senGW17}.
For example, in the case of the waveform THC0105 (see the last row of table 2) this gives
the scaling factor of $5.4/0.44 \cong 12.3$. Multiplying amplitude $h_{\rm rss}$ of 0.24 by this
factor gives $h_{\rm rss}^{50\%}$ of around 3 (in units $10^{-22}/\sqrt{\text{Hz}}$).

We then calculate $h_{\rm rss}^{50\%}$ from formula \eqref{eq:hrss} using the rescaled polarizations.
The search sensitivities $h_{\rm rss}^{50\%}$ are shown in the eighth column of table \ref{tab:senGW17}.
We also show the distance to the source corresponding to $h_{\mathrm{rss}}^{50\%}$.
We provide as a point of comparison the $h_{\rm rss}$ of the same NR waveforms used in the analysis
but assuming the distance of GW170817.  

\begin{table}
\caption{\label{tab:senGW17}
Sensitivity of the search for the waveforms consistent with the GW event GW170817.
The first column is the label of the entry from the CoRe database,
the second and the third columns are the masses of the components of the binary,
the fourth is the label of the equation of state (EOS) of the nuclear matter,
the fifth is the frequency corresponding to the maximum of the spectrum of the waveform,
the sixth is the SNR of the waveform in the O2 LIGO data assuming distance to the GW170817 merger,
the seventh column is the amplitude $h_{\rm rss}$ for the waveform obtained from formula \eqref{eq:hrss}.
The eighth column is the amplitude $h_{\rm rss}$ obtained assuming that the amplitude of the waveform is such
that its SNR gives 50\% probability of detection with $10^{-4}$ false alarm probability.
The last column is the distance corresponding to the waveform with amplitude $h_{\rm rss}^{50\%}$
taken from the previous column.}
\lineup
\begin{indented}
\item[]\begin{tabular}{@{}ccccccccc}
\br 
PM
& $m_1$
& $m_2$
&
& $f_{\rm peak}$
&
& $h_{\rm rss}$
& $h_{\rm rss}^{50\%}$
& distance
\\[0ex]
waveform
& [$M_\odot$]
& [$M_\odot$]
& EOS 
& [Hz]
& SNR
& [$10^{-22}/\sqrt{\text{Hz}}$]
& [$10^{-22}/\sqrt{\text{Hz}}$]
& [Mpc]
\\ \mr
THC0038 & 1.699 & 1.104 & BLh & 3076 & 0.22 &  0.16 & 8.1 & 0.78\\
THC0041 & 1.486 & 1.254 & DD2 & 2448 & 0.46 &  0.22 & 2.9 & 3.08 \\
THC0042 & 1.497 & 1.245 & DD2 & 2460 & 0.47 &  0.23 & 3.0 & 3.04 \\
THC0043 & 1.509 & 1.235 & DD2 & 2468 & 0.49 &  0.23 & 2.8 & 3.29 \\
THC0052 & 1.364 & 1.364 & BLh & 2992 &  0.39 &  0.22 & 3.2 & 2.77\\
THC0053 & 1.635 & 1.146 & BLh & 3104 &  0.23 &  0.17 &   5.0 &  1.32\\
THC0054 & 1.772 & 1.065 & BLh & 3088 &  0.10 &  0.12 &   7.3 &  0.66\\
THC0055 & 1.635 & 1.146 & DD2 & 2448 &  0.35 & 0.19 &  3.2 & 2.33\\
THC0063 & 1.364 & 1.364 & BLh & 3052 &  0.48 & 0.26 &  3.2 & 3.16\\
THC0064 & 1.581 & 1.184 & BLh & 2932 &  0.39 & 0.22 &  3.7 & 2.40\\
THC0065 & 1.699 & 1.104 & BLh & 3108 &  0.23 & 0.16 &  7.1 & 0.91\\
THC0066 & 1.364 & 1.364 & DD2 & 2688 &  0.41 & 0.21 &  1.8 & 4.64\\
THC0077 & 1.482 & 1.259 & BLh & 3012 &  0.49 & 0.25 &  3.9 & 2.58\\
THC0079 & 1.364 & 1.364 & LS220 & 2976 &  0.52 & 0.27 &  2.7 & 3.95\\
THC0080 & 1.364 & 1.364 & LS220 & 2960 &  0.51 & 0.26 &  3.1 & 3.46\\
THC0081 & 1.364 & 1.364 & LS220 & 2912 &  0.53 & 0.27 &  3.2 & 3.42\\
THC0082 & 1.4\0\0 & 1.33\0 & LS220 & 3024 &  0.49 & 0.26 &  3.0 & 3.54\\
THC0083 & 1.635 & 1.114 & LS220 & 3168 &   0.20 & 0.16 &  4.2 & 1.54\\
THC0087 & 1.146 & 1.635 & SFHo & 3068 &   0.33 & 0.19 &  4.6 & 1.67\\
THC0089 & 1.364 & 1.364 & SFHo & 3328 &   0.36 & 0.26 &  4.4 & 2.31\\
THC0091 & 1.452 & 1.283 & SFHo & 3360 &   0.26 & 0.21 &  4.1 & 2.03\\
THC0095 & 1.452 & 1.283 & SLy4 & 3456 &  0.30 & 0.23 &  4.7 & 1.94\\
THC0096 & 1.635 & 1.146 & SLy4 & 3000 &    0.41 & 0.22 &  3.1 & 2.78\\
THC0105 & 1.482 & 1.259 & BLh & 2968 &  0.44 & 0.24 &  3.0 & 3.25\\
\br
\end{tabular}
\end{indented}
\end{table}

\begin{figure}
\includegraphics[width=\textwidth]{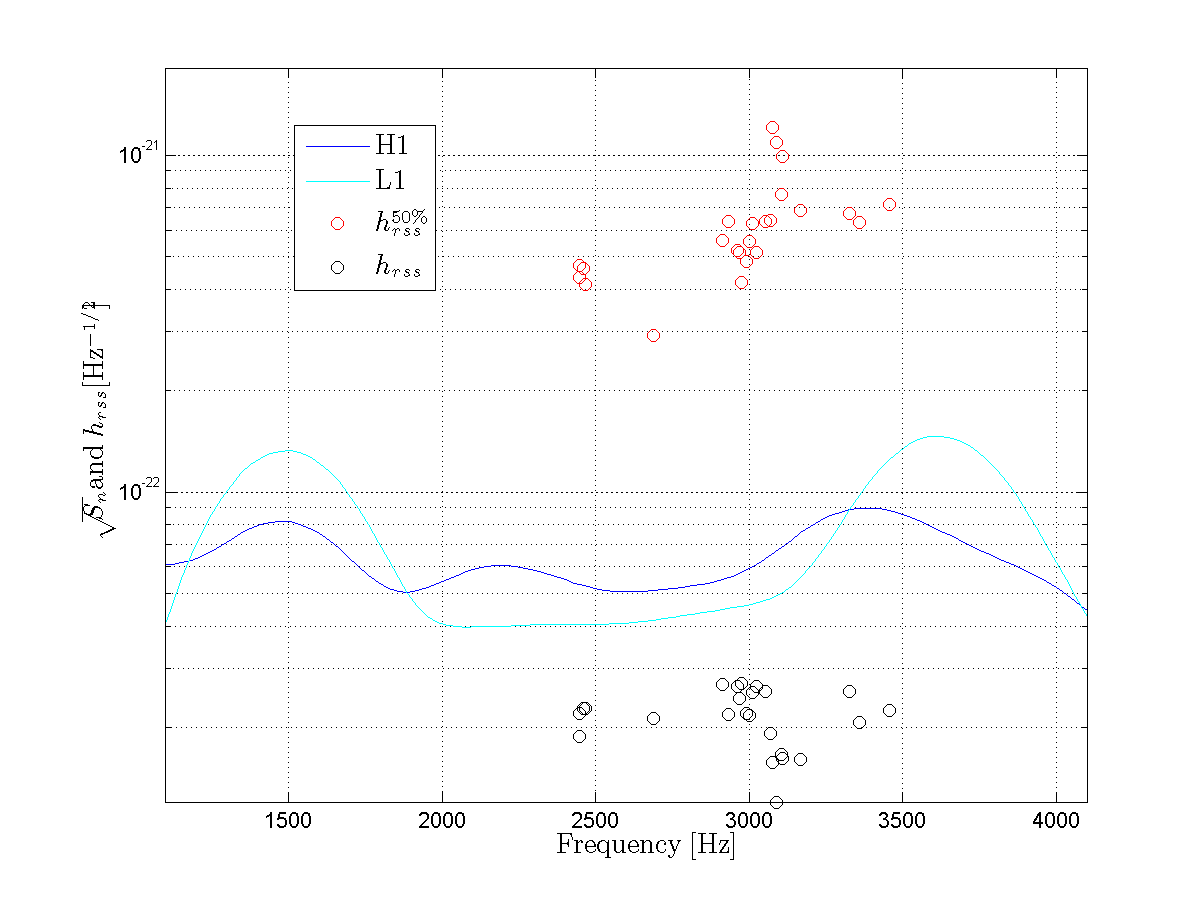}
\caption{Noise amplitude spectral density $\sqrt{S_n}$ for the two LIGO detectors (solid lines)
and detection efficiency root-sum-square strain amplitudes $h_{\text{rss}}$ at 50\% false
dismissal probability (red circles) for various postmerger waveforms.
The black circles represent the postmerger NR waveforms used in the analysis,
but at the $h_{\text{rss}}$ assuming the distance and orientation of GW170817
inferred from the premerger observation in \cite{PhysRevLett.119.161101}.}
\label{fig:sen_all}
\end{figure}

The search sensitivities are also shown in figure \ref{fig:sen_all} as functions
of the maximum frequency $f_{\text{peak}}$ of the spectrum of each waveform.
The spectral density estimates shown in figure \ref{fig:sen_all}
are spectral densities used to whiten the noises in the detectors
and to calculate SNRs of the postmerger waveforms. 

\section{Conclusions}
\label{sec:con}

We have developed a matched-filtering statistic to search for short,
lasting a few tens of milliseconds,
gravitational-wave signals following the merger of two neutron stars.
Our matched filter is an approximate model of postmerger signals
obtained through extensive numerical simulations available in the CoRe database \cite{CoRe2022}.
We have tested our method with those numerical waveforms from the CoRe database
that are consistent with the GW170817 event \cite{Nedora_2021}.
By injecting the numerically obtained waveforms into the LIGO data
we have found that with our approximate matched filter the SNR loss is
from 9\% to 55\% with respect to a perfectly matched filter.
We have also found that with our detection method,
a postmerger signal consistent with the GW170817 event
would be confidently detected if it occurred within the range (1--5)~Mpc,
depending on the numerical model. The sensitivity of our method is comparable
to the sensitivity of the search reported in \cite{Abbott_2017},
which used the coherent Wave Burst (cWB) pipeline \cite{PhysRevD.93.042004}.
We have also performed a search for a postmerger signal in the LIGO data
following the GW170817 merger using our approximate matched filter.
No significant signal was found.
In order to confidently detect short  postmerger signals
we need an increase in the sensitivity of detectors by at least one order of magnitude.
Confident detection of postmerger signals can be expected
with a planned third generation of detectors
like the Cosmic Explorer \cite{CE:2019} and the Einstein Telescope \cite{https://doi.org/10.48550/arxiv.2205.09979}.

\section*{Acknowledgments}

This research has made use of data or software obtained from the Gravitational Wave Open Science Center (gwosc.org),
a service of LIGO Laboratory, the LIGO Scientific Collaboration, the Virgo Collaboration, and KAGRA.
LIGO Laboratory and Advanced LIGO are funded by the United States National Science Foundation (NSF)
as well as the Science and Technology Facilities Council (STFC) of the United Kingdom,
the Max-Planck-Society (MPS), and the State of Niedersachsen/Germany for support of
the construction of Advanced LIGO and construction and operation of the GEO600 detector.
Additional support for Advanced LIGO was provided by the Australian Research Council.
Virgo is funded, through the European Gravitational Observatory (EGO),
by the French Centre National de Recherche Scientifique (CNRS),
the Italian Istituto Nazionale di Fisica Nucleare (INFN) and the Dutch Nikhef,
with contributions by institutions from Belgium, Germany, Greece, Hungary, Ireland, Japan, Monaco, Poland, Portugal, Spain.
KAGRA is supported by the Ministry of Education, Culture, Sports, Science and Technology (MEXT),
Japan Society for the Promotion of Science (JSPS) in Japan;
National Research Foundation (NRF) and Ministry of Science and ICT (MSIT) in Korea;
Academia Sinica (AS) and National Science and Technology Council (NSTC) in Taiwan.

We would like to thank Sebastiano Bernuzzi and Toni Font for helpful discussions.
The work was supported by the Polish National Science Centre Grant No.\ 2017/26/M/ST9/00978.

\appendix

\section{Numerical relativity postmerger waveforms}
\label{app:waveforms}

In this appendix, we explain how from different $(lm)$ modes of GW perturbations,
available in numerical relativity databases,
one can get the GW plus and cross polarization functions.
Our presentation is based on \cite{https://doi.org/10.48550/arxiv.0709.0093}.

Let $h_{ab}$ be the GW metric perturbation
in the wave zone, where $a$ and $b$ are space-time indices.
In the transverse-traceless (TT) gauge all
information about the metric perturbation is contained in the TT
tensor $h_{ij}$, where $i$ and $j$ are spatial indices. 
In the TT gauge, $h_{ij}$ can be expressed in terms of two independent
polarizations $h_+$ and $h_\times$.
The combination $h_+-ih_\times$ of the polarizations
can be decomposed into modes using spin-weighted spherical harmonics $\Ys_{lm}$ of weight $-2$:
\be
\label{eq:hDecomposition}
h_+ - ih_\times = \frac{M}{r}
\sum_{\ell=2}^{\infty}\sum_{m=-\ell}^\ell H_{\ell m}(t)\,
\Ytwo_{\ell m}(\iota,\phi),
\ee
where $r$ is the distance from the source and $M$ is the total mass of the source.
The expansion parameters $H_{lm}$ are complex functions of the retarded time $t-r$,
but if we fix $r$ to be the radius of the sphere at which  gravitational waves are extracted,
then $H_{lm}$ are functions of $t$ only.

The explicit expression for the spin-weighted spherical harmonics in
terms of the Wigner $d$-functions is
\be
\label{eq:5}
\Ys_{lm}(\iota,\phi) = (-1)^s \sqrt{\frac{2\ell+1}{4\pi}}
d^{\,\ell}_{m,s}(\iota) e^{im\phi},
\ee
where
\begin{align}
\label{eq:6}
d^{\,\ell}_{m,s}(\iota) &= \sum_{k = k_1}^{k_2}
\frac{(-1)^k[(\ell+m)!(\ell-m)!(\ell+s)!(\ell-s)!]^{1/2}}
{(\ell +m-k)!(\ell-s-k)!k!(k+s-m)!}
\nonumber\\[1ex]&\quad
\times \left(\cos\left(\frac{\iota}{2}\right)\right)^{2\ell+m-s-2k}
\left(\sin\left(\frac{\iota}{2}\right)\right)^{2k+s-m},
\end{align}
with $k_1 = \textrm{max}(0, m-s)$ and $k_2=\textrm{min}(\ell+m,\ell-s)$.
For reference,
\begin{subequations}
\label{eq:7}
\begin{align}
\Ytwo_{22}(\iota,\phi) &= \sqrt{\frac{5}{64\pi}}(1+\cos\iota)^2e^{2i\phi} \,,
\\[1ex]
\Ytwo_{21}(\iota,\phi) &= \sqrt{\frac{5}{16\pi}}  \sin\iota( 1 + \cos\iota )e^{i\phi} \,,
\\[1ex]
\Ytwo_{20}(\iota,\phi) &= \sqrt{\frac{15}{32\pi}} \sin^2\iota \,,
\\[1ex]
\Ytwo_{2-1}(\iota,\phi) &= \sqrt{\frac{5}{16\pi}}  \sin\iota( 1 - \cos\iota)e^{-i\phi} \,,
\\[1ex]
\Ytwo_{2-2}(\iota,\phi) &= \sqrt{\frac{5}{64\pi}}(1-\cos\iota)^2e^{-2i\phi}\,.
\end{align}
\end{subequations}

The modes $h_+^{(\ell m)}$ and $h_\times^{(\ell m)}$ of the two polarizations of the signal
are defined through the relation
\be
\label{eq:11}
h_+^{(\ell m)}(t) - i h_\times^{(\ell m)}(t) := \frac{M}{r}H_{\ell m}(t)\,.
\ee
The modes $h_{+,\times}^{(\ell m)}$ are provided
in the CoRe database of numerical waveforms \cite{CoRe2022}.
In the database also the modes' amplitudes $A^{(\ell m)}$ and phases $\Phi^{(\ell m)}$
are provided, in terms of which the $(\ell m)$ modes are given by
\begin{subequations}
\label{eq:rec}
\begin{align}
h_+^{(\ell m)}(t) &= A^{(\ell m)}(t) \cos\big(\Phi^{(\ell m)}(t)\big) \,,
\\[1ex]
h_\times^{(\ell m)}(t) &= -A^{(\ell m)}(t) \sin\big(\Phi^{(\ell m)}(t)\big) \,.
\end{align}
\end{subequations}

For the case of the $l = 2$ mode,
assuming equatorial symmetry, the two polarization functions are given by
\begin{subequations}
\label{eq:hpc}
\begin{align}
h_+(t;\iota,\phi) &= \Re \sum_{m=-2}^2 \big(h_+^{(2m)}(t)- i h_\times^{2m}(t)\big)\,\Ytwo_{2m}(\iota,\phi)
\nonumber\\[1ex]
&= \sqrt{\frac{5}{64\pi}}
\Big( 2 (1+\cos^2\iota)\big( h_+^{(22)}(t)\cos2\phi + h_\times^{(22)}(t)\sin2\phi \big)
\nonumber\\[1ex]&\qquad\qquad
+ 4\sin\iota \big( h_+^{(21)}(t)\cos\phi + h_\times^{(21)}(t) \sin\phi \big) + 
\sqrt{6}\sin^2\iota \, h_+^{(20)}(t) \Big),
\\[2ex]
h_\times(t;\iota,\phi) &= -\Im \sum_{m=-2}^2 \big(h_+^{(2m)}(t)-i h_\times^{2m}(t)\big)\,\Ytwo_{2m}(\iota,\phi)
\nonumber\\[1ex]
&= \sqrt{\frac{5}{64\pi}}
\Big( 4\cos\iota\, \big(h_\times^{(22)}(t)\cos2\phi - h_+^{(22)}(t)\sin2\phi\big)
\nonumber\\[1ex]&\qquad\qquad
+ 2 \sin2\iota\, \big(h_\times^{(21)}(t)\cos\phi - h_+^{(21)}(t)\sin\phi\big) +
\sqrt{6}\sin^2\iota \, h_\times^{(2 0)} \Big).
\end{align}
\end{subequations}
Let us introduce the following modified mode functions:
\begin{subequations}
\label{eq:recmod}
\begin{align}
H_+^{(\ell m)}(t;\phi) &:= A^{(\ell m)} \cos(\Phi^{(\ell m)}(t) + m\phi),
\\[1ex]
H_\times^{(\ell m)}(t;\phi) &:= -A^{(\ell m)} \sin(\Phi^{(\ell m)}(t) + m\phi).
\end{align}
\end{subequations}
Using these functions the two polarizations $h_+$ and $h_\times$
can be expressed in a more compact form as
\begin{subequations}
\label{eq:hpcmod}
\begin{align}
h_+(t;\iota,\phi) &= \sqrt{\frac{5}{64\pi}} \Big( 2 (1+\cos^2\iota) \, H_+^{(2 2)}(t;\phi)
+ 4 \sin\iota \,  H_+^{(2 1)}(t;\phi) + \sqrt{6}\sin^2\iota \, H_+^{(2 0)}(t;\phi) \Big),
\\[1ex]
h_\times(t;\iota,\phi) &= \sqrt{\frac{5}{64\pi}} \Big( 4 \cos\iota \, H_\times^{(2 2)}(t;\phi)
+ 2\sin2\iota\, H_\times^{(2 1)}(t;\phi) +  \sqrt{6}\sin^2\iota \, H_\times^{(2 0)}(t;\phi) \Big).
\end{align}
\end{subequations}
Thus we see that the angle $\phi$ can be absorbed into the phases $\Phi^{(\ell m)}(t)$.

As the $(l=2,m=\pm2)$ mode is dominant,
to a very good approximation we can neglect in Eqs.\ \eqref{eq:hpcmod}
contributions due to the modes (2,1) and (2,0).
Then we have:
\begin{subequations}
\label{eq:hpcmod2}
\begin{align}
\label{eq:hpcmod2p}
h_+(t;\iota,\phi) &\cong \sqrt{\frac{5}{4\pi}}\frac{1+\cos^2\iota}{2}\,H_+^{(22)}(t;\phi)
\nonumber\\[1ex]
&= \sqrt{\frac{5}{4\pi}}\frac{1+\cos^2\iota}{2}\,A^{(22)}(t)\cos\big(\Phi^{(22)}(t) + 2\phi\big),
\\[2ex]
\label{eq:hpcmod2c}
h_\times(t;\iota,\phi) &\cong \sqrt{\frac{5}{4\pi}}\cos\iota\,H_\times^{(22)}(t;\phi)
\nonumber\\[1ex]
&= -\sqrt{\frac{5}{4\pi}}\cos\iota\,A^{(22)}(t)\sin\big(\Phi^{(22)}(t) + 2\phi\big).
\end{align}
\end{subequations}

Let us now assume that the time decay of the amplitude $A^{(22)}(t)$
is simply modelled as $A^{(22)}(t)=A_0e^{-ot}$ (where $A_0$ and $o$ are positive constants),
and that the phase $\Phi^{(22)}(t)$ can be approximated as a quadratic function of time $t$,
$\Phi^{(22)}(t)=2\pi f t + 2\pi \gamma t^2$ (where $f$ and $\gamma$ are another constants).
Then, after introducing new constants
\be
A_{0+}:=\sqrt{\frac{5}{4\pi}}A_0\frac{1+\cos^2\iota}{2},
\quad A_{0\times}:=\sqrt{\frac{5}{4\pi}}A_0\cos\iota,
\quad \beta:= \frac{\pi}{2} + 2\phi,
\ee
equations \eqref{eq:hpcmod2} can be rewritten in the form of Eqs.\ \eqref{hphc}
(where we have replaced $\Phi^{(22)}$ just by $\Phi$).

\section{Signal-to-noise ratio, Fisher matrix, false alarm probability}
\label{app:FAP}

In this Appendix, we shall present an approximate calculation of the accuracy of estimation of the parameters
with our match-filtering method and an estimate of the false alarm probability (FAP) for our search.
The basic tools are the optimal SNR $\rho$ and the Fisher information matrix $\Gamma_{ij}$,
which for a given signal $s$ are defined as
\begin{align}
\label{eq:snr_opt}
\rho &:= \sqrt{(s|s)},
\\[1ex]
\label{eq:FM_gen}
\Gamma_{ij} &:= \Big(\frac{\partial s}{\partial \theta_i}\Big|\frac{\partial s}{\partial \theta_j}\Big).
\end{align}
Assuming that over the bandwidth of the signal the spectral density of the noise is nearly constant,
the scalar product $(x|y)$ is approximately given by
\be
\label{eq:sp}
(x|y) \cong \frac{2}{S_c}\int_0^\To x(t) y(t)\,\mathrm{d}t,
\ee
where $S_c$ is the one-sided spectral density at the frequency
for which the power spectrum of the signal is maximum.
The SNR and the Fisher matrix are then given by
\begin{align}
\rho^2 &\cong \frac{2}{S_c}\int_0^\To s^2(t,\btheta)\,\md t,
\\[1ex]
\label{eq:FM_app}
\Gamma_{ij} &\cong \frac{2}{S_c}\int_0^\To
\frac{\partial s(t,\btheta)}{\partial \theta_i}
\frac{\partial s(t,\btheta)}{\partial \theta_j}\,\md t,
\end{align}
where the signal $s(t,\btheta)$ is defined in Eqs.\ \eqref{RF7}--\eqref{RF8} and \eqref{Phi}
[we assume that $s(t,\btheta)=0$ for $t < 0$ and for $t > \To$].
The vector $\btheta$ collects five parameters of the signal:
$\btheta = (A_c, A_s, \tau, \omega, \omega_1)$,
where $\tau=1/o$, $\omega = 2\pi f$, and $\omega_1 = 2 \pi \gamma$.

We calculate the SNR ratio and the Fisher matrix numerically assuming the signal is discretely sampled.
Then the scalar product \eqref{eq:sp} is given by
\be
\label{eq:spd}
(x|y) \cong \frac{1}{\sigma^2}\sum_{t=1}^N x(t) y(t),
\ee
where $\sigma^2$ is the variance of the noise and $N$ is the number of the signal's samples.
We have used the relation
\be
S_o = 2\sigma^2 \Delta t,
\ee
where $\Delta t$ is the sampling time period.

The general formula for the number of cells in the parameter space reads
(see section 6.2 of \cite{JK:2009})
\be
\label{eq:NC}
N_c = \frac{\gamma(m/2+1)}{(\pi/2)^{m/2}}
\int_V\sqrt{\det \tilde{\Gamma}}\, \md V,
\ee
where $\gamma$ is the Euler's gamma function,
$m$ is the number of intrinsic parameters,
$V$ is the (hyper)volume of the intrinsic parameters space,
and $\tilde{\Gamma}$ is the \emph{reduced} Fisher matrix computed for the intrinsic parameters
[in our case $m=3$ and the intrinsic parameters are $(\tau,\omega,\omega_1)$].
The number of cells $N_c$ defines the number of independent realizations
of our matched-filtering statistic ${\mathcal P}$ in the parameter space that we search.
The false alarm probability $P^T_F$ for the threshold ${\mathcal P}_0$ of the ${\mathcal P}$-statistic
is then given by
\be
\label{FP}
P^T_F(\mathcal{P}_0)
= 1 - \big[1 - P_F(\mathcal{P}_0)\big]^{N_c},
\ee
where $P_F$ is the probability distribution of the ${\mathcal P}$-statistic when
there is no signal. For Gaussian noise $2{\mathcal P}$ has a central $\chi^2$
distribution with 2 degrees of freedom. 

\begin{figure}
\includegraphics[width=0.75\textwidth]{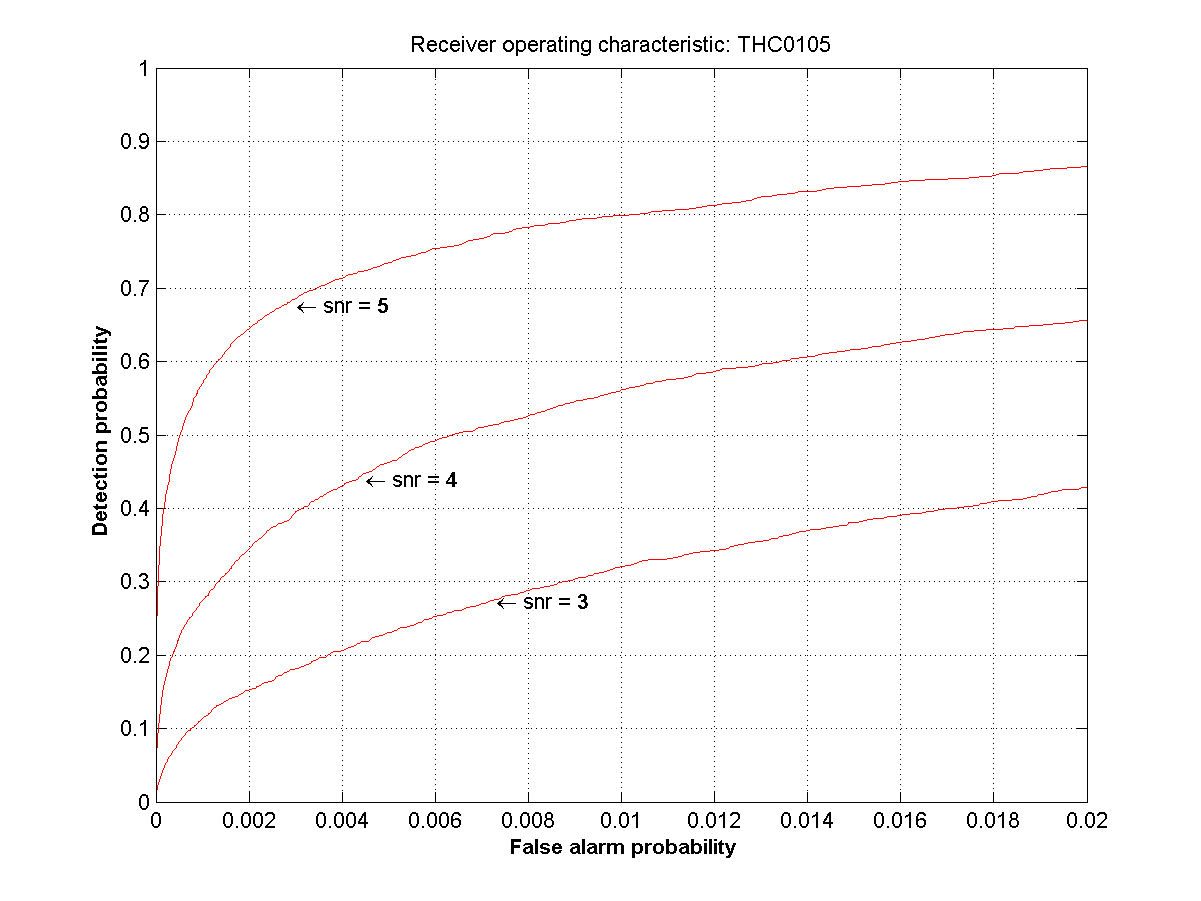}
\caption{Receiver operating characteristic (ROC) for the detection of the postmerger waveform THC0105
in the network of two LIGO detector data. The ROC curves are parametrized by the injected network SNR.}
\label{fig:ROC_data}
\end{figure}

With the above formula for the false alarm probability,
we can simulate the receiver operating characteristic (ROC) for our statistic.
This is presented in figure \ref{fig:ROC_data} for the case
of the postmerger waveform THC0105 from the CoRe database.
The ROC curves are parametrized by the optimal SNR $\rho$.
To obtain a ROC curve for a given $\rho=\rho_0$ we first calculate,
by numerically inverting the formula \eqref{FP},
the threshold $\mathcal{P}_0(\alpha;\rho_0)$ as a function of the false alarm probability $\alpha$.
Then we inject to the network of LIGO detectors data,
waveforms THC0105 with the SNR $\rho_0$ (in the same way
as for our Monte Carlo simulations presented in section \ref{sec:injections}).
Then we count how many threshold crossings of $\mathcal{P}_0(\alpha;\rho_0)$ we have
from the 1000 signals that we inject for each $\alpha$.
This gives as the probability of detection $P_d(\mathcal{P}_0(\alpha;\rho_0))$
as a function of $\alpha$ for a specific SNR equal to $\rho_0$.

\end{document}